\documentclass[aps,prd,amsfonts,amssymb, amsmath, showpacs, showkeys, a4paper, nofootinbib, superscriptaddress, 11pt, raggedbottom]{revtex4-2}
\usepackage{graphicx, subfig}

\usepackage{notoccite}
\usepackage{amsmath, latexsym, amssymb}
\usepackage{geometry}
\usepackage{hyperref}
\usepackage{braket}
\usepackage{mathtools}
\usepackage{color}
\usepackage{float}
\usepackage{ulem}

 \usepackage{float}
\usepackage{babel}
\usepackage{enumitem}

\DeclareMathOperator{\sech}{sech}

\begin{document}

\title{Impact of cavities on the detection of  quadratically coupled ultra-light dark matter}
\author{Clare Burrage}
\email{clare.burrage@nottingham.ac.uk}
\author{Angus Macdonald}
\email{angus.macdonald@nottingham.ac.uk}
\affiliation{School of Physics and Astronomy, University of Nottingham,
University Park, Nottingham NG7 2RD, UK}
\author{Michael P. Ross}
\email{mpross2@uw.edu}
\author{Gray Rybka}
\email{grybka@uw.edu}
\affiliation{University of Washington, Seattle, WA 98195, USA}
\author{Elisa Todarello}
\email{elisa.todarello@berkeley.edu}
\affiliation{School of Physics and Astronomy, University of Nottingham,
University Park, Nottingham NG7 2RD, UK}

\begin{abstract}
    Ultra-light scalar fields may explain the nature of the dark matter in our universe. If such scalars couple quadratically to particles of the Standard Model the scalar acquires an effective mass which depends on the local matter energy density. The changing mass causes the field to deviate from its cosmological value in experimental environments. In this work we show that the presence of a local over-density enclosing the experiment, for example a cavity, vacuum chamber, or satellite can strongly suppress the value of the scalar and its gradient in the interior. This makes detection of such scalar dark matter challenging, and significantly relaxes constraints on strongly coupled  models. 
    We also discuss the possibility that quadratically coupled ultra-light scalar dark matter could be detected by the differential measurement of the force on two cavities of the same mass but different internal structure. 
\end{abstract}
\maketitle
\section{Introduction}
Galactic and cosmological scale observations, particularly in the face of the robustness of General Relativity \cite{Will:2014kxa}, have motivated the existence of an unobserved ``dark matter'' constituting around $25\%$ of the energy balance of the Universe \cite{Planck:2018vyg}. Light scalars have been considered as an origin of this phenomenon \cite{Frieman:1995pm,Hui:2021tkt, Antypas:2022asj}: a class of feebly interacting particle, with mass $\ll1\mbox{ eV}$.
Such scalars are common objects in models of  physics beyond the Standard Model, appearing in theories beyond General Relativity \cite{Burrage:2018dvt,Clifton:2011jh,Joyce:2014kja}, and  models attempting to resolve fine tuning in the Standard Model such as the QCD axion \cite{Weinberg:1977ma, Wilczek:1977pj}. 

Light scalars must be produced non-thermally in the early universe to be oscillating in their potential today, and thus behave as cold dark matter. Examples of such production mechanisms are vacuum realignment~\cite{Abbott:1982af, Dine:1982ah, Preskill:1982cy, Arias:2012az}, decay of heavy fields such as moduli~\cite{Broeckel:2021dpz} or decay of topological defects~\cite{Davis:1986xc}.  If the field mass is $
\ll1$~eV, typical occupation numbers are so high that it is possible to model the dark matter in our Galaxy as an incoherent superposition of classical scalar waves~\cite{Foster:2017hbq}\footnote{See Refs~\cite{Sikivie:2016enz, Chakrabarty:2017fkd, Dvali:2017ruz, Eberhardt:2022rcp} for a discussion of quantum corrections to the classical field behaviour.}.
In order to detect or constrain such scalar dark matter fields, it is necessary to  understand how these particles interact with matter. Absent a principle to forbid such interactions, we expect them to be present, and they may be universal, for example: arising from a conformal rescaling of the metric, they may be mediated through a Higgs portal coupling, or arise directly from interactions between the scalar and fundamental particles.  Key work by Damour and Donoghue, \cite{Damour:2010rp, Damour:2010rm}, showed that if such a light scalar evolves with time, it induces time dependent shifts in effective constants of physics - particularly in the fine structure constant, fermionic masses and the QCD confinement scale, and how the couplings to fundamental particles can be expressed as couplings to atoms. This framework has been used to predict scalar-mediated phenomena such as `$5^{\rm th}$ forces', Equivalence Principle violating effects, and shifts to atomic spectra. These have in turn been used to justify searches for such particles through means such as tests of the universality of free fall, and frequency shifts in atomic clocks \cite{Kim:2022ype, Panda:2023nir, Hees:2018fpg, Filzinger:2023qqh, Arvanitaki:2016fyj, Sponar:2020gfr, Elder:2025tue, Vermeulen:2021epa, Berge:2017ovy}.  Such theories will also be constrained with future experimental advances, including with atom interferometry \cite{MAGIS-100:2021etm,Bertoldi:2021rqk,AEDGE:2019nxb,Abend:2023jxv,Badurina:2019hst}, hyperfine transitions \cite{Hees:2016gop,Kennedy:2020bac},  molecular spectroscopy \cite{Oswald:2021vtc} and nuclear clocks \cite{Banks:2024sli}.

To date, much of this work has focussed on scalars that couple linearly to standard model fields. More recently however, there has been a growing interest in studying quadratically coupled scalars, where a symmetry forbids a linear coupling, so the quadratic coupling controls the leading order behaviour. Quadratic couplings are expected to appear in models for axions and axion-like particles \cite{Bauer:2023czj,Grossman:2025cov},  symmetric Higgs-portal scalars \cite{OConnell:2006rsp,Patt:2006fw} or those with mirror symmetry \cite{Delaunay:2025pho}. Generalisation of the principles of Refs.~\cite{Damour:2010rp, Damour:2010rm} to the quadratic case, has allowed constraints on such models to be derived from laboratory and satellite measurements \cite{Hees:2018fpg, Bauer:2024hfv, VanTilburg:2024xib,Gue:2025nxq,Gan:2025nlu,Gue:2025nxq}. 

A key feature of the phenomenology of quadratically coupled scalars, which is not present for linearly coupled models,  is that they may be screened or enhanced in the vicinity of massive objects \cite{Banerjee:2022sqg}, depending on the sign of the coupling term. For positive couplings, when an object becomes sufficiently compact we find that adding more mass no longer causes a proportional response of the scalar field.  For negative couplings the scalar field can be significantly enhanced around certain massive objects (although this can be disrupted by propagating waves of the scalar field \cite{Banerjee:2025dlo,delCastillo:2025rbr})  Some impacts of this have been understood, and the effects included in existing constraints  \cite{Hees:2018fpg,Bauer:2024yow}. For example, for experiments conducted in the vicinity of the Earth it is important to determine whether the Earth enhances or suppresses the scalar field.  
However, 
effects deriving from interaction between scalars and laboratory environments, and the geometry of test masses, often treated as point particles,  have been less studied.
In this article, we aim to demonstrate the need to consider such factors. 

In this work, we will examine the behaviour of quadratically coupled scalars in and around enclosed spaces, or cavities, which we use to give an approximate description of experimental apparatus and test objects with non-trivial geometry.
In  Sec.~\ref{sec:Profiles}, we introduce quadratically coupled scalar dark matter models. Then  Sec.~\ref{sec:scalarprofiles}  discusses  the scalar field profile around massive objects, including solid spheres, and  spherical and cylindrical cavities, which we then use as an approximation to explore how scalar-induced phenomena may be altered in an experiment conducted in an enclosed laboratory environment. 
 Finally in Sec.~\ref{sec:Forces}, we shift focus back to the spherical cavity case, and look at the $5^{\rm th}$ forces acting on the cavities themselves due to the enhancement or suppression of the dark matter around the cavities or the Earth. We determine the equivalence principle violations that come from cavities of the same mass and dimension but different internal structure. 
 We use this section to discuss the importance of test mass geometry on the magnitude of such forces. Expressions throughout this paper will employ natural units: $\hbar=c=1$.

\section{Quadratically coupled scalar dark matter} \label{sec:Profiles}
In this work we  consider the coupling of a canonical, massive  scalar field to a local classical non-relativistic matter density, $\rho(\vec{x})$, through the action:
\begin{equation}
  S=\int d^4x\left(\frac{1}{2}(\partial\phi)^2-\frac{1}{2}m^2\phi^2-\frac{1}{2}\rho\alpha\phi^2\right)\;.
  \label{eq:scalar_field_action}
\end{equation}
The constant $\alpha$, has dimensions $[\mbox{mass}]^{-2}$, and parametrises the  interaction strength between scalar and matter. When matter is non-relativistic a coupling of the scalar field to energy density can arise from a universal coupling to matter, such as a conformal coupling \cite{Burrage:2018dvt,Clifton:2011jh,Joyce:2014kja}, or through a coupling of the scalar to the Higgs boson ~\cite{Binoth:1996au,Schabinger:2005ei,Patt:2006fw,Ahlers:2008qc,Batell:2009yf,Englert:2019eyl,Englert:2020gcp, Brax:2023udt}, which when the Higgs is integrated out leaves an effective coupling of the form in equation \eqref{eq:scalar_field_action} at low energies. It can also arise from coupling the scalar directly to the fundamental bosons and gluons \cite{Hook:2017psm,Kim:2023pvt,Bauer:2024hfv,Bauer:2023czj,Bauer:2024yow,Grossman:2025cov, Chattopadhyay:2024rha}, in which case  composition dependence may arise, discussed further in section \ref{sec:Dilaton_Coefficients}.  In this work we are interested in the leading order effects of a universal coupling to matter. If the coupling to matter is composition dependent, the effects discussed in this work might vary somewhat in strength for different materials, but will otherwise remain. We assume that the $\rho \alpha \phi^2$ is the leading order interaction between the scalar field and matter, but allow for the possibility that higher order interactions may be present.  To ensure that such higher order terms remain subdominant we will require $|\alpha\phi^2|\ll1$.
\\\\
From Eq.~\eqref{eq:scalar_field_action}, it is simple to derive the equation of motion for the scalar field:
\begin{equation}
    \Box\phi=-\left(m^2+\rho\alpha\right)\phi\;.
    \label{eq:equation_of_motion}
\end{equation}
From equation \eqref{eq:equation_of_motion}, it is apparent that a local energy density can shift the effective (squared) mass of the scalar field,  $m_{eff}^2 = m^2+\rho\alpha$. We note that when  $\rho\alpha<-m^2$  the effective squared mass of a field oscillating about $\phi=0$ may be negative. As will be discussed in later sections, this may cause the expectation value of the field expectation to roll to a new minimum, which can have important implications with regards to the stability of field solutions, and the validity of the scalar theory in Eq.~\eqref{eq:scalar_field_action} \cite{Banerjee:2025dlo,Hook:2017psm,  Balkin:2020dsr, Balkin:2022qer, Balkin:2023xtr, Gomez-Banon:2024oux}. For the largest values of $\alpha$ that we will consider, everyday materials can lead to large increases in the  mass of a scalar that is otherwise extremely light in vacuum, leading to Compton wavelengths on the order of the size of experimental apparatus:
\begin{equation}
 \lambda_C=   \frac{h}{m_{\rm eff}c} \approx 10^{-5} \left(\frac{\mbox{GeV}^{-2}}{\alpha}\right)^{1/2} \left(\frac{\mbox{g/cm}^3}{\rho}\right)^{1/2}\mbox{cm}\;,
\end{equation}
where we have reinstated Planck's constant, $h$, and the speed of light, $c$. The implications of this for the possibility of detecting the dark matter scalar are the subject of this article. 

\subsection{Connection to `dilaton coefficients'}
\label{sec:Dilaton_Coefficients}
Previous work reporting constraints on quadratically coupled ultra-light dark matter, has often done so in terms of `dilaton coefficients' controlling the interaction of the scalar with different fundamental particles, e.g. Ref.~\cite{Hees:2018fpg}. For such a canonical massive scalar the interaction Lagrangian is written as:
\begin{equation}
    \mathcal{L}_{int} = \frac{(\kappa\phi)^2}{2}\left[\frac{d_e}{4}F_{\mu\nu}F^{\mu\nu}-\frac{d_g\beta_3}{2g_3}F^A_{\mu\nu}F^{A~{\mu\nu}}-\sum_{i={u,d,e}}(d_{m_i}+\gamma_{m_i}d_g)m_i\bar{\psi}_i{\psi}_i\right] \;,
\end{equation}
with $\kappa=\sqrt{4\pi G}$. $F$ and $F^A$ are photon and gluon field strength tensors respectively, $g_3$ is the QCD gauge coupling, which runs according to $\beta_3$, $\gamma_{m_i}$ provides the anomalous dimension responsible for the running of quark masses, and $d_e,~d_g, ~d_{m_i}$ are the dimensionless `dilaton coefficients'. These interactions induce an effective scalar dependence in the fundamental constants of nature - namely the fine structure constant, fermionic masses, and the QCD mass scale:
\begin{equation}
\begin{split}
    \alpha_{EM}(\phi) &= \alpha_{EM}\left(1+d_e\frac{(\kappa\phi)^2}{2}\right)\;,\\
    m_i(\phi) &= m_i\left(1+d_{m_i}\frac{(\kappa\phi)^2}{2}\right)\;,\\
    \Lambda_3(\phi) &=\Lambda_3\left(1+d_g\frac{(\kappa\phi)^2}{2}\right)\;.
\end{split}
\label{eq:Fundamental_Shifts}
\end{equation}
From this dependence, Damour and Donoghue determine the effective $\phi$-dependent shift to the masses of atoms $m_A$ in Refs.~\cite{Damour:2010rp, Damour:2010rm}:
\begin{equation}
m_A(\phi)=m_A\left(1+\tilde{\alpha}_A\frac{(\kappa\phi)^2}{2}\right)\;,
\label{eq:Scalar_Shifted_Atomic_Mass}
\end{equation}
where $\tilde{\alpha}_A$ is expressed in terms of the $d_i$ parameters:
\begin{equation}
    \tilde{\alpha}_A =  d_g+Q_{\hat{m}}(d_{\hat{m}}-d_g)+Q_{\delta m}(d_{\delta m}-d_g)+Q_{m_e}(d_{m_e}-d_g)+Q_{e}d_{e} \;.
    \label{eq:Alpha_Definition}
\end{equation}
The dimensionless ``dilaton charges'' $Q_i$ may be calculated semi-empirically in terms of the proton number, mass number, and mass of the atoms to which the field is coupled. For their value for commonly used materials we refer the reader to Ref.~\cite{Hees:2018fpg}.
The constants $d_{\hat{m}}$ and $d_{\delta m}$ are defined as
\begin{equation}
\begin{split}
    d_{\hat{m}} &= \frac{ m_{u} d_{m_u} + m_{d} d_{m_d}}{m_{u} + m_{d}}\;,\\
    d_{\delta m} &= \frac{ m_{u} d_{m_u} - m_{d} d_{m_d}}{m_{u} - m_{d}}\;.
\end{split}
\end{equation}

\noindent If all matter had the same composition, our universal coupling $\alpha$ in Eq. \eqref{eq:scalar_field_action} would be defined as  
\begin{equation}
    \alpha = \kappa^2\tilde{\alpha}_A \;.
    \label{eq:Alpha_Conversion}
\end{equation}
In the rest of this work we are concerned with the effects of the presence of different matter distributions, and their shape, on the dark matter scalar field profile, we  will ignore composition dependant effects, except when 
discussing existing constraints in the literature.

\subsection{Homogeneous Background}
The vacuum solutions for Eq.~\eqref{eq:equation_of_motion}, will be  the  boundary conditions for the field far from the massive objects we will consider. Specifically, throughout this paper, we will make the assumption the field is isotropic and homogeneous far from any matter. This assumption is made for simplicity, though incorporating a stochastic, non-isotropic dark matter background may have phenomenological implications \cite{Banerjee:2025dlo,delCastillo:2025rbr,  Foster:2017hbq}. Relaxing this assumption is beyond the scope of this article. Under these circumstances,  the field loses its spatial dependence, such that equations of motion simply reduce to:
\begin{equation}
    \partial_t^2\phi=-m^2\phi
    \label{eq:background_equations_of_motion}\;,
\end{equation}
from which the homogeneous  vacuum solution is:
\begin{equation}
    \phi_H(t)=\phi_\infty\cos(mt+\delta) \;,
    \label{eq:Homogenous_Solution}
\end{equation}
where $\delta$ is some unknown phase factor. 
The local dark matter density is then:
\begin{equation}
    \rho_{DM}=\frac{1}{2}m^2\phi_\infty^2~.
    \label{eq:homogeneous_amplitude}
\end{equation}
Throughout this work, we will take the local dark matter energy density today to be $\rho_{DM}\simeq0.4~\mbox{GeV}~\mbox{cm}^{-3}$ \cite{deSalas:2020hbh}. 

\section{Scalar field profiles in the presence of matter}\label{sec:scalarprofiles}
In this Section we show how the presence of compact objects gives rise to spatially varying scalar field profiles. Starting with the  known solution around spheres of constant density, and then considering spherical and cylindrical cavities. 

\subsection{Spherically Symmetric Uniform Object} \label{sec:sphere}
\begin{figure}
    \centering
    \includegraphics[width=0.5\linewidth]{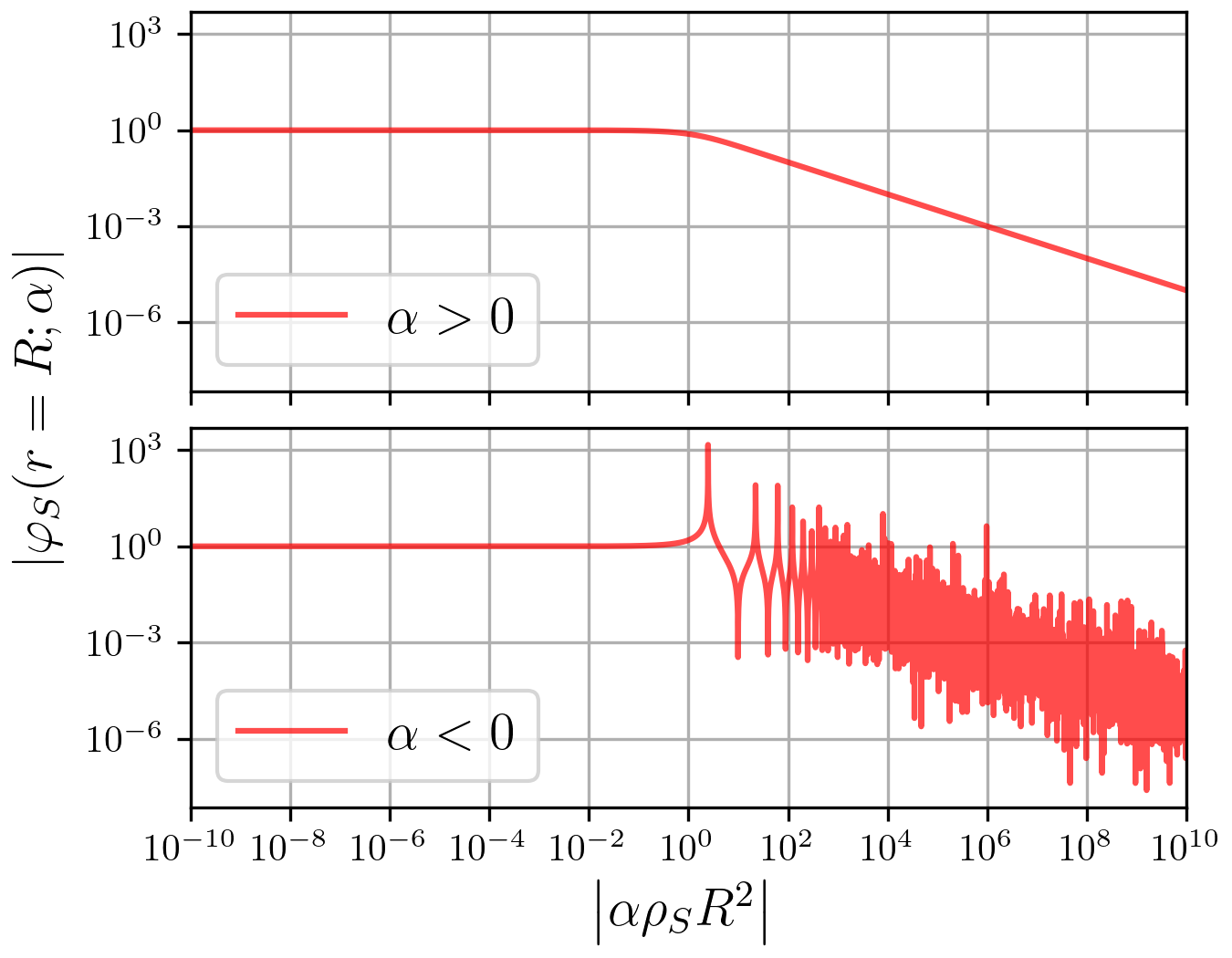}
    \caption{Evaluation of the field profile at the surface of a solid sphere radius $R$, varying $|\alpha \rho _SR^2|=|kR|^2$. The top panel illustrates the case $\alpha>0$, in which the field at the surface is screened for $\alpha=1/\rho R^2$. The lower panel demonstrates the case $\alpha<0$. Maxima and minima indicate the locations of field divergences and zeros respectively. We note that the magnitude of the field is displayed given that, in the $\alpha<0$ case, sign changes occur between adjacent extrema. }
    \label{fig:Sphere_Surface_Value}
\end{figure}

\begin{figure}
    \centering
    \includegraphics[width=0.45\linewidth]{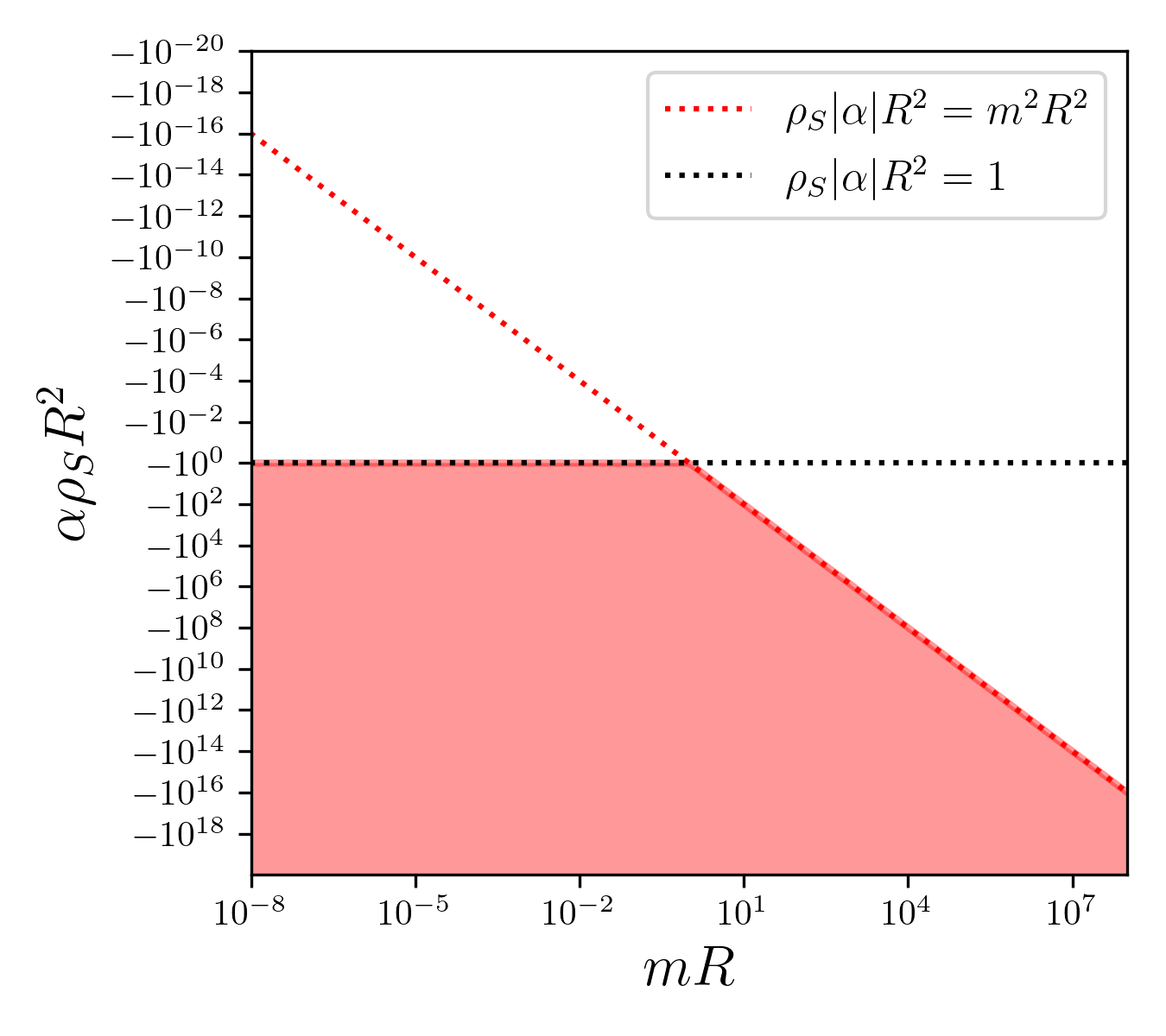}
    \caption{The region of instability of the scalar field solutions around a uniform sphere. The mass $m$ and coupling strength $\alpha$ of the scalar field are appropriately scaled according to the sphere's density $\rho_S$ and radius $R$. The \textit{red dotted line} illustrates the boundary below which the field is tachyonic. The \textit{black dotted line} illustrates the boundary under which the wavelength of the field inside matter is sufficiently small to allow its expectation value to roll away from zero. The \textit{red shaded region} illustrates the intersection of these regions: $\alpha<-\max{(\frac{m^2}{\rho}, \frac{1}{\rho R^2})}$, whereby the field solutions in Eq.~\eqref{eq:Solution_Uniform_Sphere} are expected to be unstable to perturbations.}
    \label{fig:regions_of_validity}
\end{figure}

The simplest object we can consider is a spherical body with uniform density and composition. 
We recall the solutions for the field sourced by a massive sphere \cite{Hees:2018fpg}, of radius $R$, and density:
\begin{equation}
\rho(\vec{x\,})=\rho_S\Theta(R-r) \; .
\end{equation}
for constant $\rho_S$.
We assume this sphere is isolated, such that at an infinite distance from the object, the field tends to the limit of the of the homogeneous background. This allows us to  factorise the field into homogeneous and spatially varying components:
\begin{equation}
    \phi=\phi_H(t)\varphi_S(r),
\end{equation}
 such that $\varphi_S$ obeys the boundary condition:
\begin{equation}
    \lim_{r\rightarrow\infty}\varphi_S(r)=1,
    \label{eq:Homogeneous_Condition}
\end{equation}
Inserting this ansatz into Eq.~\eqref{eq:equation_of_motion} gives the equation of motion for $\varphi_S$:
\begin{equation}
    \nabla^2\varphi_S(r)=\alpha\rho(r)\varphi_S(r)~.
\end{equation}
Spherical symmetry makes this a simple problem to solve, and ultimately gives a solution of the form:
\begin{equation}
    \varphi_S(r)=\begin{cases}
        A^{in}_S\frac{\cosh(kr)}{r}\ + B^{in}_S\frac{\sinh(kr)}{r} & r<R\\
        A^{out}_S + \dfrac{B^{out}_S}{r} & r>R~.\\
    \end{cases}
\end{equation}
Here, $k$ acts as growth rate/wavenumber for the field inside the sphere, given as $k=\sqrt{\rho_S\alpha}$. Note that when $\alpha$ is positive: $k\in\mathbb{R}$ and
the field solutions are hyperbolic, but when  $\alpha$ is negative:  $k\in i\mathbb{R}$ and the field solutions oscillate spatially inside the sphere. The factors $A^{in/out}_S,\; B^{in/out}_S$ are merely constants of integration, which may be determined by requiring that the field be continuous and differentiable at $r\in\{0,~R\}$, and by imposing the boundary condition in Eq.~\eqref{eq:Homogeneous_Condition}. The solution is then:
\begin{equation}
    \varphi_S(r)=\begin{cases}
        \sech(kR)\frac{\sinh(kr)}{kr} &r<R\\
        1+\dfrac{q(kR)R}{r} & r>R \;,
    \end{cases}
    \label{eq:Solution_Uniform_Sphere}
\end{equation}
where we define
\begin{equation}
    q(kR)=\frac{\tanh(kR)-kR}{kR} ~.
    \label{eq:q_function}
\end{equation}
This solution may separated into two regimes based on the behaviour of $q(kR)$:
\begin{equation}
    q(k R )\approx \begin{cases}
        -\frac{1}{3}(kR)^2 & |kR| \ll 1 \\
        -1 & |kR| \gg 1\;.
    \end{cases}
\end{equation}
The case $|kR|\ll 1$ is interpreted as a regime of weak coupling between sphere and field, in which the characteristic length scale of the field in matter, $1/|k|$, is considerably larger than the sphere. Hence, in this region, the field is minimally perturbed, and $\varphi_S \approx1$. Conversely, in the case $|kR|\gg 1$, the sphere and field can be considered strongly coupled. As can be seen in  Fig.~\eqref{fig:Sphere_Surface_Value}, this leads to a general trend in which the field is increasingly screened in the vicinity of the sphere as $|kR|$ increases.
\\\\
However, in the case $\alpha<0$, we observe additional phenomenology. Since $k\in i\mathbb{R}$, the $\tanh(kR)$ term in Eq.~\eqref{eq:q_function} now becomes $\sim\tan(|k|R)$. The field therefore diverges asymptotically at $|k| R=\pi\left(n+\frac{1}{2}\right)$, for integer $n$, and hence has an enhanced amplitude in the neighbourhood of these points.
Whilst the condition $|\alpha\phi^2|\ll1$ does not hold sufficiently close to these divergences, there may still be $|k|$ for which some degree of enhancement can be physical, depending on the value of $m$ associated with the field. Conversely, we note that there are additionally values for which $\tan(|k|R)=|k|R$, and so the effective scalar charge of the source vanishes, $q(kR)=0$, and the field profile around the sphere will be indistinguishable from the uncoupled case.
\\\\
Additionally, for the case $\alpha<0$, it is necessary to acknowledge the restricted parameter space in which the solutions in Eq.\eqref{eq:Solution_Uniform_Sphere} are stable. These solutions assume the field to have an expectation value $\left<\phi\right>=0$, both inside and outside matter. This is clearly a reasonable assumption if the mass-squared of the scalar is positive within the sphere: $m_{eff}^2=m^2+\rho\alpha>0$. Less trivially, it is also a valid in the case where the field is tachyonic inside the sphere: $m_{eff}^2=m^2+\rho\alpha<0$, but where $|k|R\ll1$. Under these circumstances, the field expectation value has insufficient space inside the sphere to roll to a new minimum. In the case that both of these conditions are violated: $m^2\lesssim\rho|\alpha|$ and $|k|R\gtrsim1$, we expect the field solutions in Eq.~\eqref{eq:Solution_Uniform_Sphere} to be unstable to perturbations. The value at which the field stabilises will be determined by terms beyond the quadratic cutoff of the field theory in Eq.~\eqref{eq:scalar_field_action}. As such, we are unable to describe this region of parameter space - shown in Fig.~\ref{fig:regions_of_validity} -  in our scalar theory.
\\\\
In our analysis, we will use this profile as a simplified description of the field profile sourced by the Earth, defining:
\begin{equation}
    \varphi_\oplus(r) = \varphi_S(r\;;R_\oplus,~k_\oplus) \; ,  
\end{equation}
for which $k_\oplus = \sqrt{\rho_\oplus\alpha}$, $\rho_\oplus\simeq5510 ~\mbox{kg m}^{-3}$ and $R_\oplus = 6.4\times10^{6}~m$.

\subsection{Spherical Cavity}\label{sec:Spherical_Cavity_Profile}
As described above, the amplitude of quadratically coupled scalars can be both enhanced and suppressed around matter. We are interested in how these effects may manifest in the context of enclosed environments. To this end, we consider the field around a spherical shell or ``cavity'' with inner radius $R_1$, outer radius $R_2$ and density $\rho(r)$:
\begin{equation}
    \rho(r)=\begin{cases}
        0 & r<R_1\\
        \rho_c & R_1<r<R_2\\
        0 & r>R_2\;\;\;\;\;\;\;\;\;\;\;\;\;\;.
    \end{cases} 
\end{equation}
where $\rho_c$ is the constant density of the cavity walls. This may be considered an approximate description of apparatus, such as optical cavities, vacuum chambers, or even satellites, in which scalar-field searches may be housed. Similarly, to the uniform sphere case, we consider the decomposition of the field into a homogeneous and spatially varying component:
\begin{equation}
    \phi=\phi_H(t)\varphi_c(\vec{x\,})~.
    \label{eq:Homogeneous_Condition_Cavity}
\end{equation}
As in the previous sub-section we could solve for the field profile around this cavity on a homogeneous dark matter background.  However, we choose to include here the possibility that this cavity sits on a background scalar profile, sourced by some other object (e.g. the Earth). 
As such, we choose to assume that boundary conditions to the field sourced by the cavity field include a simple anisotropic, linear component. 
This allows us to show that the results we derive are not specific to an isolated object in a homogeneous background. Additionally, it allows us to predict how field gradients are modified in the presence of a cavity. This is necessary for the calculation of scalar-mediated $5^{\rm th}$ forces, both for particles inside the void of the cavity, and on cavities themselves as test masses, which we will consider in Sec.~\ref{sec:Forces}. As such, we consider the following boundary conditions:
\begin{equation}
    \lim_{r\rightarrow\infty}\varphi_c(r,\theta)=a_0+a_1r\cos\theta.
    \label{eq:Cavity_Infinity_Boundary_Condition}
\end{equation}
where, $r,~\theta$ form part of the spherical co-ordinate basis about the cavity centre, defining $\theta\in \{0\leq\theta<\pi\}$ such that $\theta=0$ aligns with the direction of the field gradient. A caveat here is that, under our own description, the background field to the cavity should tend to the homogeneous solution away from the matter sourcing it. This requirement is in conflict with Eq.~\eqref{eq:Cavity_Infinity_Boundary_Condition}, which diverges as  $r\rightarrow\infty$. However, we may discard this problem if the background field is sufficiently slowly varying, such that it can be described only by the leading order terms in its Taylor expansion in the vicinity of the cavity. These terms can be used to define $a_0,~a_1$. For example, by  expanding $\varphi_\oplus(r'; k_\oplus,R_\oplus)$, about the position of the cavity relative to the centre of the Earth, $\vec{r}_c$, we might define $a_0,~a_1$ using the field sourced by the Earth:
\begin{equation}
\begin{split}
    a_0+a_1r\cos\theta=\left[1 + q(k_\oplus R_\oplus )\frac{R_\oplus }{r_c}\right] -\left[q(k_\oplus R_\oplus )\frac{R_\oplus }{r_c^2}\right]r\cos\theta\;.
\end{split}
\label{eq:Earth_Taylor_Expansion}
\end{equation}
The validity of this interpretation when evaluating $a_0, ~a_1$ for different parameters is discussed in Sec.~\ref{sec:Forces}. With the inclusion of these boundary conditions, the system is no longer  spherically symmetric and  the equation of motion for the field is:
\begin{equation}
    \frac{1}{r^2}\partial_r(r^2\partial_r\varphi_c) + \frac{1}{r^2\sin\theta}\partial_\theta\left(\sin\theta\partial_\theta\varphi_c\right)=\rho\alpha\varphi_c\;.
    \label{eq:Cavity_Spatial_Equation_Of_Motion}
\end{equation}
By imposing the boundary conditions in  Eq.~\eqref{eq:Cavity_Infinity_Boundary_Condition}, demanding continuity and differentiability  of the field at the boundaries $r\in \{0, ~ R_1,~R_2\}$, and through orthogonality of angular modes, solutions simplify to the form: 
\begin{equation}
        \varphi_c(r) = \begin{cases}
        A_0^{in} + A_1^{in}r\cos\theta  & r<R_1\\
        \\
        \begin{split}
            & \left(A_0^{walls}[j_0(-ikr)] + B_0^{walls}[(i) y_{0}(-ikr)]\right) \\
            &+ \left(A_1^{walls}[(-i)j_1(-ikr)] + B_1^{walls}[(-1) y_{1}(-ikr)]\right)\cos\theta
        \end{split} & R_1<r<R_2\\
        \\
        \left(a_0 + B_0^{out}r^{-1}\right) +\left(a_1r+B_1^{out}r^{-2}\right)\cos\theta & r>R_2\;,
    \end{cases}
    \label{eq:Solution_Cavity}
\end{equation}
where $A^{in/out}_l,~B^{in/out}_l$ are constants of integration, $j_l(x),~y_l(x)$ are spherical Bessel functions of the first and second kind respectively, $k=\sqrt{\alpha\rho_c}$ and $i$ is the imaginary unit. Further detail on the origin of these terms, and analytical expressions for these coefficients are are presented in Appendix \ref{app:Coeffs}. 
\\\\
We note that, as with the field solutions around a uniform sphere, there should be a region of parameter space similar to that in Fig.~\ref{fig:regions_of_validity}, under which the solutions defined by Eq.~\eqref{eq:Solution_Cavity} are unstable to perturbations. In this work, the scale and density of the cavities considered will be smaller than the corresponding properties of Earth - which is modelled as a uniform sphere in Sec.~\ref{sec:sphere}. As such, the region of instability for the field solutions around these cavities will be contained within the region of instability for field solutions around the Earth. For simplicity, we always choose to exclude the full region in which the field solutions around the Earth are unstable in our analyses.
\subsubsection*{The Cavity Interior Field}
We now draw particular attention to the field inside the cavity, which contains only a linear and constant term - reflecting the boundary conditions used. From the analytical form of these terms, we find that each is modified from the linear background field solution in the absence of the cavity by a multiplicative factor:
\begin{align}
       a_0\rightarrow A^{in}_0&=a_0\times\frac{1}{kR_1\sinh(k(R_2-R_1))+\cosh(k(R_2-R_1))}\label{eq:Cavity_Interior_Coefficients_A0}\;,\\
        a_1\rightarrow A^{in}_1&=a_1\times\frac{3 kR_2}{3\sinh(k(R_2-R_1))+3kR_1\cosh(k(R_2-R_1))+(kR_1)^2\sinh(k(R_2-R_1))}\label{eq:Cavity_Interior_Coefficients_A1}\;.
\end{align}\\
\begin{figure}
    \centering
    \includegraphics[width=0.7\linewidth]{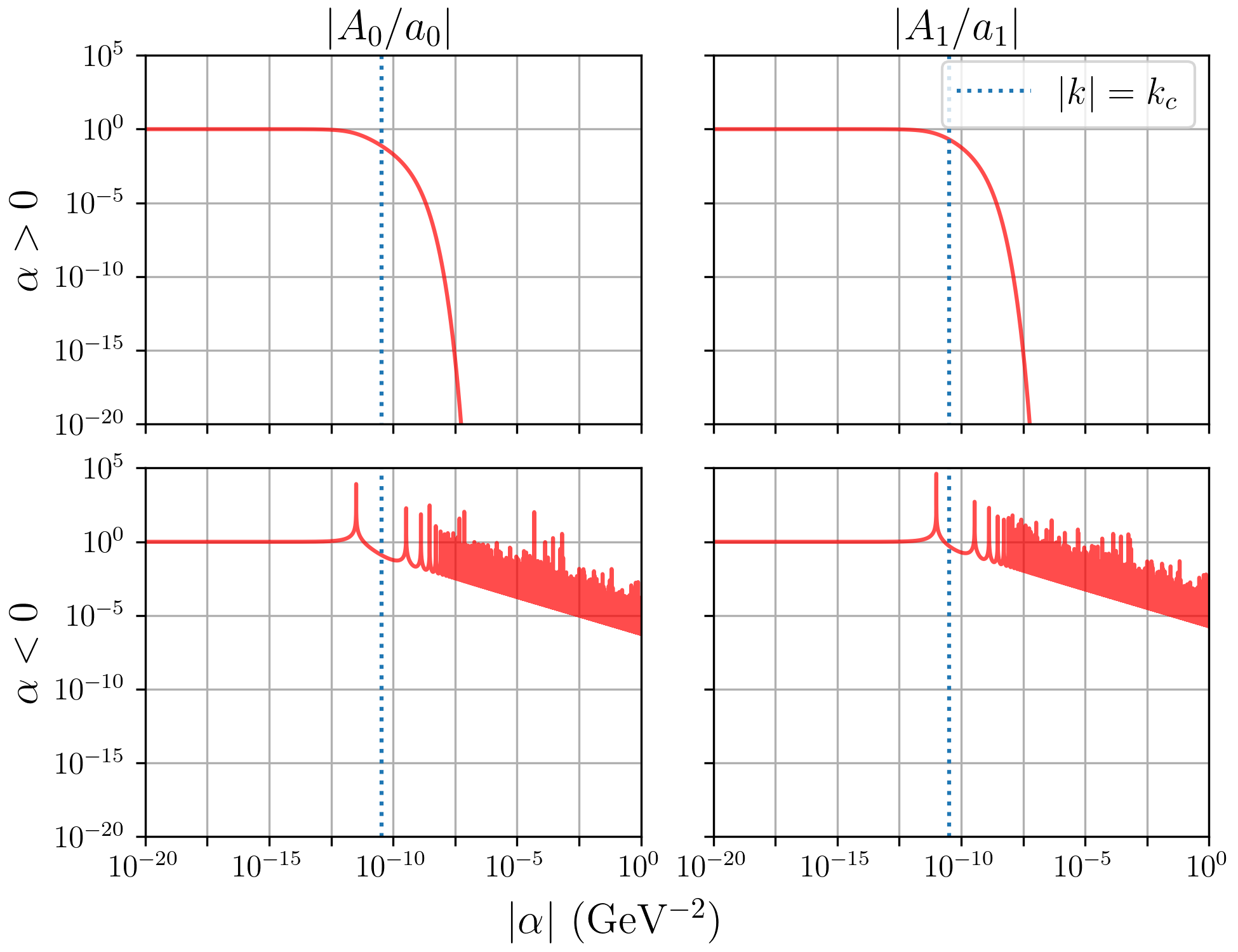}
    \caption{The modification of coefficients describing the field inside a spherical cavity: $R_1=10~\mbox{cm}$, $R_2=11~\mbox{cm}$ and $\rho_c=2700~\mbox{kg m}^{-3}$. Panels in the top row relate to coefficients for the positive $\alpha$ case, and the bottom row the negative $\alpha$ case. The left panels give the magnitude of the ratio between the constant component of the field profile inside the cavity, as in \eqref{eq:Solution_Cavity}, and the boundary condition in Eq.~\eqref{eq:Cavity_Infinity_Boundary_Condition}. The right gives the same but for the linear component of the profile and boundary conditions. Note that the magnitude is plotted since for the $\alpha<0$ case, components of the field profile inside the cavity may change sign with respect to the boundary conditions. The blue dashed line indicates where $\alpha = k_c^2/\rho_c$.}
    \label{fig:Spherical_Cavity_Interior_Coefficients}
\end{figure}
We demonstrate the effect of this modification in Fig. \ref{fig:Spherical_Cavity_Interior_Coefficients} for a small test cavity, inner radius $R_1=10~\mbox{cm}$ and outer radius $R_2=11~\mbox{cm}$, and with density similar to aluminium $\rho_{c}=2700~\mbox{kg m}^{-3}$. These dimensions have been chosen as to approximate the scale and properties of laboratory scale apparatus such as optical cavities. Other choices of size and material will exhibit qualitatively similar behaviour. 
\\\\
As with the uniform spherical case, we may choose to broadly divide the solution for the field into two regimes, by comparing the  value of $|k|$ to the thickness of the wall of the cavity. The critical value for a cavity is: $k_c=1/\left(R_2-R_1\right)$. 
For $|\alpha|\ll k_c^2/\rho_c$, such that the wavelength of the field in the cavity walls is larger than the cavity thickness $R_2-R_1$, Fig. \ref{fig:Spherical_Cavity_Interior_Coefficients} shows that the cavity has minimal impact on the amplitude of the field in the interior. Indeed, Eqs. \eqref{eq:Cavity_Interior_Coefficients_A0}, \eqref{eq:Cavity_Interior_Coefficients_A1} demonstrate that $A_0,~ A_1$ should be well approximated by $a_0,~ a_1$ for $|k|\ll k_c$, aside for the region $1/R_1\lesssim|k|<k_c$, in which polynomial decay arising from the $kR_1$ terms in the denominators could be observed.
\\\\
Where $|k|>k_c$, the cavity can have a significant effect on the scalar field profile.
In the case $\alpha>0$ we observe that both the constant and linear components of the field appear to be rapidly suppressed as $\alpha$ grows. This is corroborated by algebraic expressions for these coefficients in Eqs. \eqref{eq:Cavity_Interior_Coefficients_A0}, \eqref{eq:Cavity_Interior_Coefficients_A1}, the denominators of which grow exponentially with $k/k_c$. 
For the case $\alpha<0$, the behaviours of $A_0$ and $A_1$ are less simply characterised. Generally, they still follow a decaying trend, though this is power law decay rather than exponential suppression, and thus considerably slower than the $\alpha>0$ case. This is a consequence of the hyperbolic terms in Eqs. \eqref{eq:Cavity_Interior_Coefficients_A0}, \eqref{eq:Cavity_Interior_Coefficients_A1} becoming oscillatory for negative $\alpha$.
Additionally, for negative $\alpha$ there are values of $k$ for which the  denominators of the coefficients $A_0,~A_1$ are zero, leading to values of $\alpha$ for which the field inside the cavity is enhanced above the background, as can be seen in  Fig. \ref{fig:Spherical_Cavity_Interior_Coefficients}. From Eqs. \eqref{eq:Cavity_Interior_Coefficients_A0}, \eqref{eq:Cavity_Interior_Coefficients_A1}, we see $A_0$ is explicitly singular where
\begin{equation}
  \tan(|k|(R_2-R_1))=\frac{1}{|k|R_1}  ~,
\end{equation}
and likewise $A_1$ is singular where
\begin{equation}
  \tan(|k|(R_2-R_1))=\frac{3|k|R_1}{(|k|R_1)^2-3} ~.  
\end{equation}
Whilst we expect enhancement to be physical around these points, caution should be taken since the necessary condition $|\alpha\phi^2|\ll1$ may not hold for all $m,~\alpha$ in their neighbourhoods. It should be noted that Fig. \ref{fig:Spherical_Cavity_Interior_Coefficients} does not exactly show the number, nor the singular nature of these divergences due to limited resolution. 
Importantly, as cavity wall density, and wall thickness are increased, the threshold value $|\alpha|=k_c^2/\rho_c$ decreases, and hence cavity effects on the field become more significant at weaker couplings.
\\\\
The scalar field behaviour described above may have important implications for current constraints on quadratically coupled scalars from laboratory and satellite  experiments, which may be approximated as contained within a spherical cavity. These experiments search for evidence of scalar induced $5^{\rm th}$ forces, or other phenomena from shifts to fundamental constants (Sec.~\ref{sec:Dilaton_Coefficients}) which, for a quadratically coupled scalar, will scale as $\nabla\phi^2 \sim A_0A_1$ or $\phi^2\sim A_0^2$.
In the $\alpha>0$ case, if $\alpha\gg{k_c^2}/{\rho_c}$, then these phenomena will be exponentially suppressed. Hence, if there exists a quadratically coupled dark matter scalar  with $\alpha$ in this regime, it would be effectively absent in this environment, decoupling the dark matter from ordinary matter and making a detection of dark matter induced phenomena almost impossible. In the $\alpha<0$ case, the general power-law decay of the cavity's interior field would have a similar, albeit less substantial, suppressing effect on detectable phenomena. On the other hand, the divergences in the internal field will have the opposite effect, enhancing the field.

\subsection{Cylindrical Cavity}\label{sec:Cylinder}
\begin{figure}[th]
    \centering
    \includegraphics[width=0.8\linewidth]{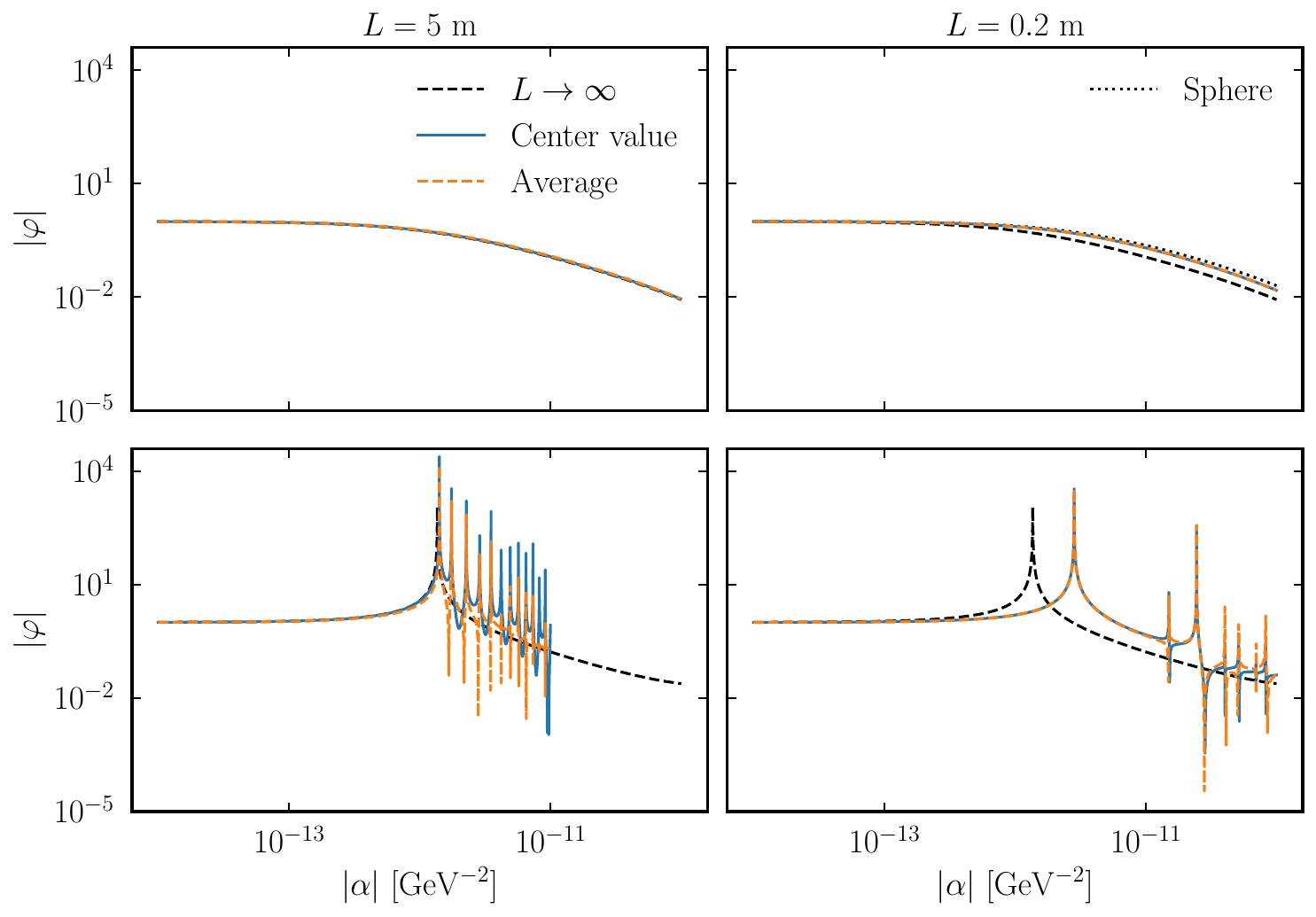}
    \caption{Absolute value of $\varphi$ at the centre (blue) of an aluminium cylindrical cavity with inner radius $R=0.1$~m and wall thickness $\delta R = \delta L = 1$~cm. The average field value inside the cavity is shown in orange. The top row corresponds to $\alpha >0$, while the bottom row to $\alpha <0$. The black dashed (dotted) lines mark the analytical result for an infinite cylinder (sphere).
    }
    \label{fig:center_values_cyl}
\end{figure}
To determine whether the suppression and enhancement found in the previous section occur for cavities that are not spherically symmetric we solve numerically the equation
\begin{equation}
    \nabla^2\varphi(r, z) = \alpha \rho(r,z) \varphi(r,z) \enspace,\label{eq:lapl_cyl}
\end{equation}
where $r$ is now the cylindrical radial coordinate, and $\rho$ the matter density of a cylindrical aluminium cavity.
We consider two cavities of different length, both with an inner radius $R=0.1$~m and a wall thickness $\delta R = \delta L = 1$~cm. The first cavity has an inner length $L=5$~m. We choose such a large length to test our numerical results against the analytic ones for an infinitely long cavity, which can be found in Appendix~\ref{app:infinite_cyle}. The second cavity has a length $L=0.2$~m, and is thus closer to the spherical geometry discussed in the previous Sections. With these choices, we can bracket the behaviour of cylindrical cavities with different $R$ to $L$ ratios.

We solve Eq.~\eqref{eq:lapl_cyl} using the Finite Element Method, with the implementation of Ref.~\cite{skfem2020}. We choose a simulation box of size $10 L \times 10 R$ and test that increasing the box size does not affect our results.
We solve imposing the Neumann boundary condition $\partial\varphi(r, z)/\partial r = 0$ at $r=0$, and the
Dirichlet boundary condition $\varphi(r, z) = 1$ on the remaining boundaries. We recursively refine our mesh in the region near the cavity to accurately resolve field oscillations.

Figure~\ref{fig:center_values_cyl} shows the field value at the centre of the cavity and its average inside the cavity as a function of the coupling $\alpha$. On the top row, we see that for positive $\alpha$ the average field value in the cavity is essentially identical to the value at the centre, both approaching zero for increasing $\alpha$. The top left panel shows excellent agreement between the numerical results for the 5~m long cavity and the analytical result for infinite $L$. As expected, the behaviour of the short cylindrical cavity falls between that of the long cylinder and sphere.

On the bottom panels of Fig.~\ref{fig:center_values_cyl}, we  see the behaviour for negative coupling. As in the spherical case, the field value tends to decrease as $|\alpha|$ increases, except at specific values, for which the field is enhanced. Every spike in the blue curve, showing the value of the field at the centre of the cavity, corresponds to an abrupt switch in the sign of $\varphi$.

\begin{figure}[t]
    \centering
    \includegraphics[width=1\linewidth]{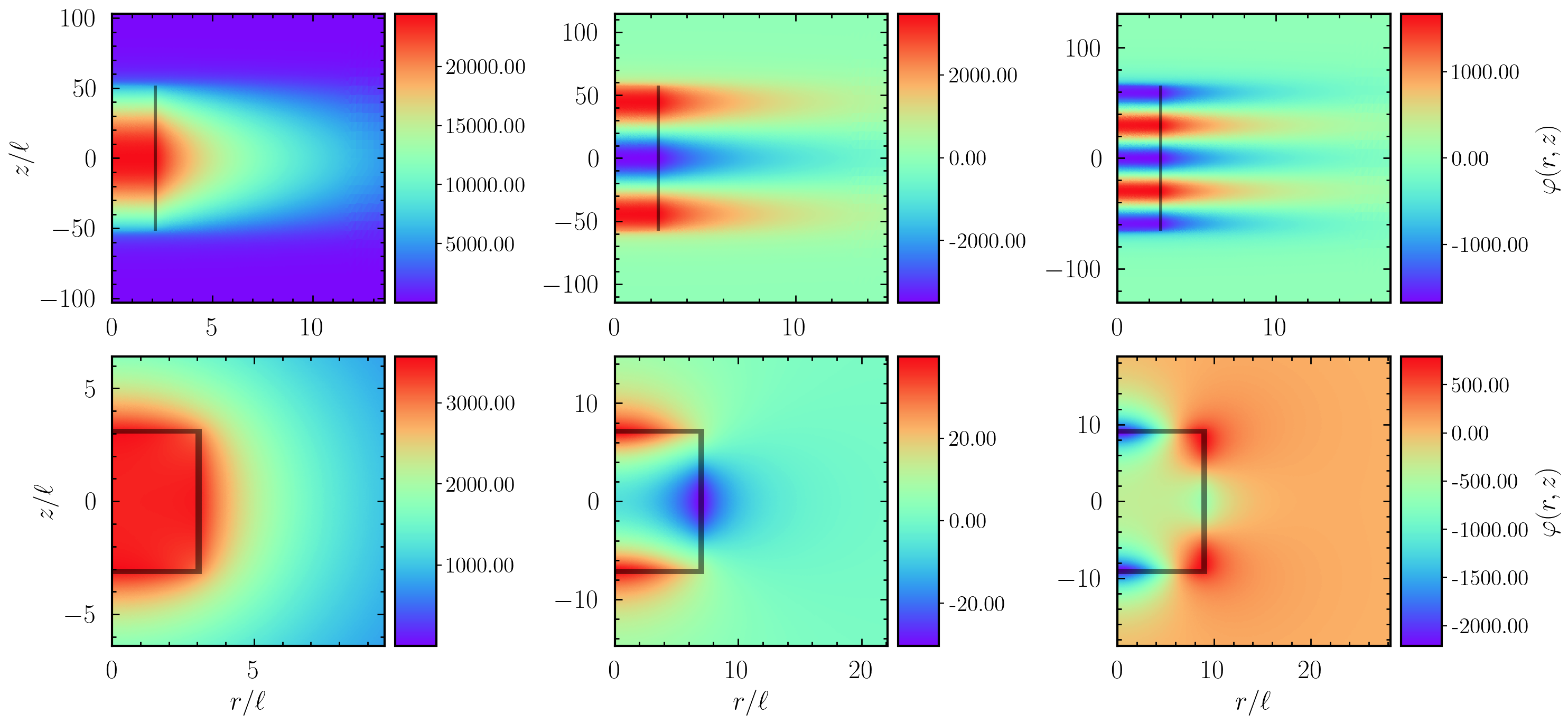}
    \caption{Field profile for $\alpha$ corresponding to the first three peaks, $\alpha = (1.41, 1.75,2.27) \times 10^{-12}~\mathrm{GeV}^{-2}$ for the long (top), and $\alpha = (2.83, 15.0, 24.5) \times 10^{-12}~\mathrm{GeV}^{-2}$ for the short (bottom) cavities. Notice that the range of field values differs in each panel. The black line shows the position of the cavity walls. 
    }
    \label{fig:field_cyl}
\end{figure}

Figure~\ref{fig:field_cyl} shows the solution for the long (top) and short (bottom) cavities, for the lowest three values of $\alpha$ corresponding peaks of Fig.~\ref{fig:center_values_cyl}. Lengths are expressed in terms of the characteristic length $\ell = 1/ \sqrt{|\alpha|\rho}$.
In the long cavity case, the additional peaks compared to the $L\to\infty$ case appear, when the $L$ is an odd multiple of the field half-wavelength in the $z$-direction. We observe a different behaviour for the short cavity, for which the oscillations are confined in the neighbourhood of the walls and seem to have a lower amplitude than in the long cavity case.

\subsection{Impact on Existing Constraints}
\label{sec:Existing_Constraints}
\begin{figure}\centering
\subfloat{\includegraphics[width=.45\linewidth]{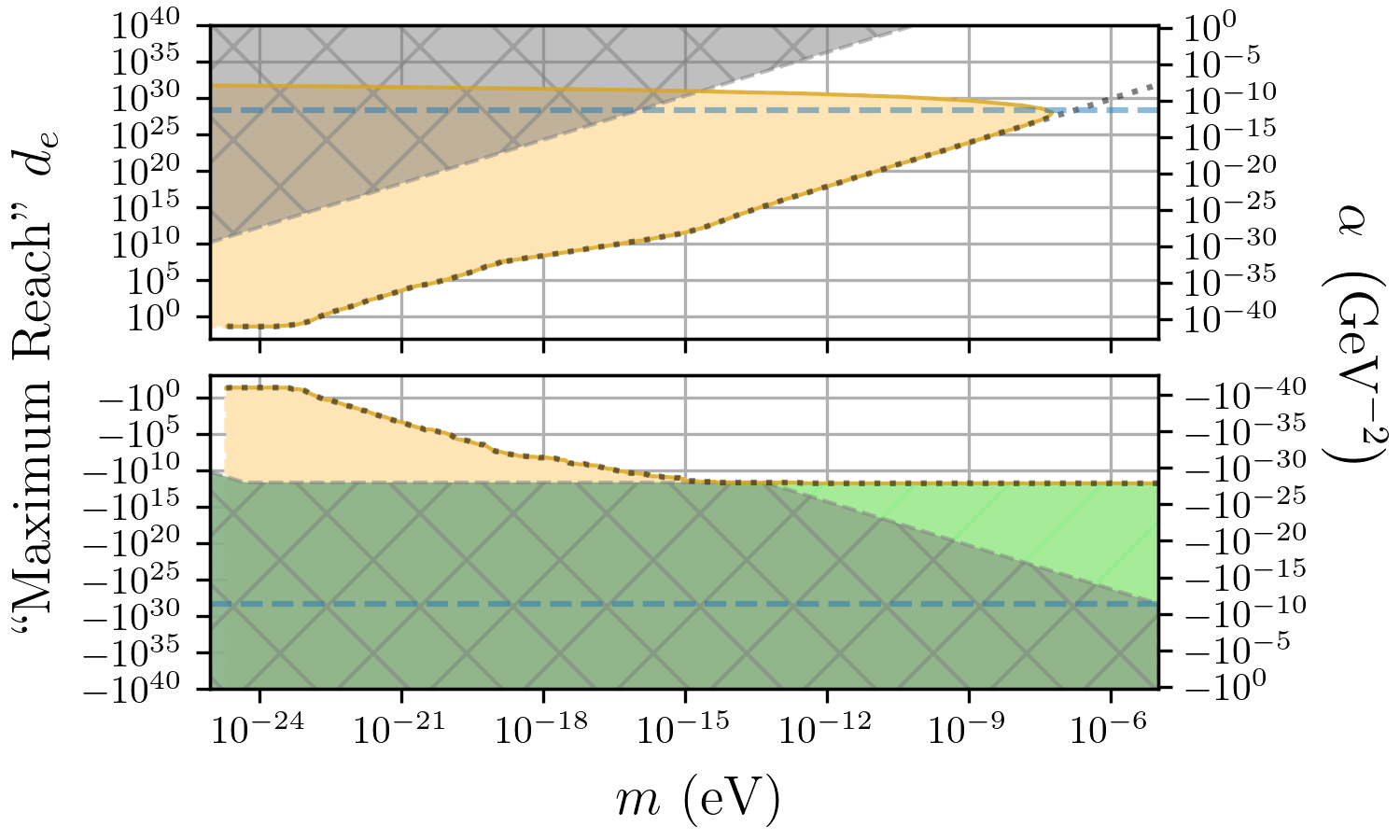}}\hfill
\subfloat{\includegraphics[width=.45\linewidth]{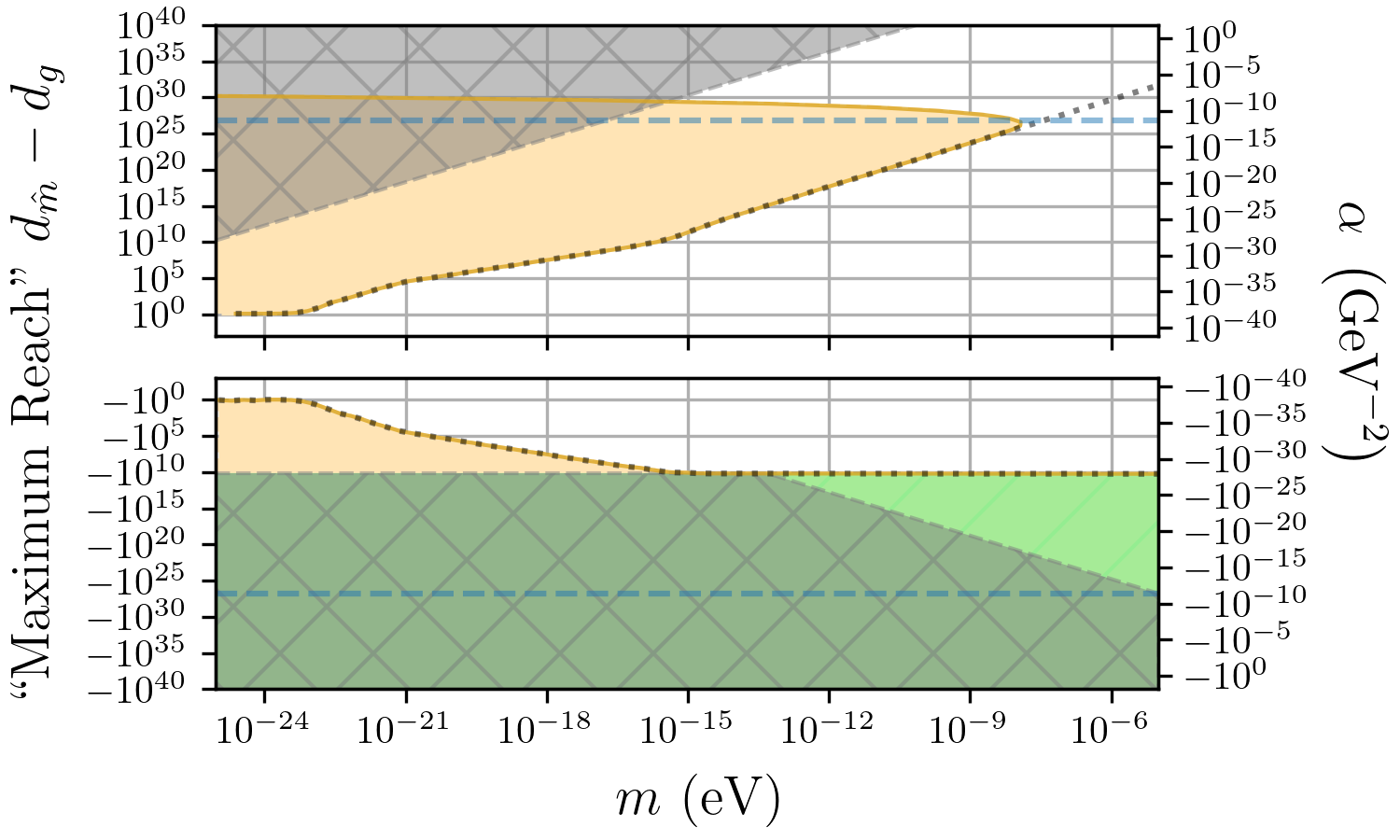}}\par 
\subfloat{\includegraphics[width=.45\linewidth]{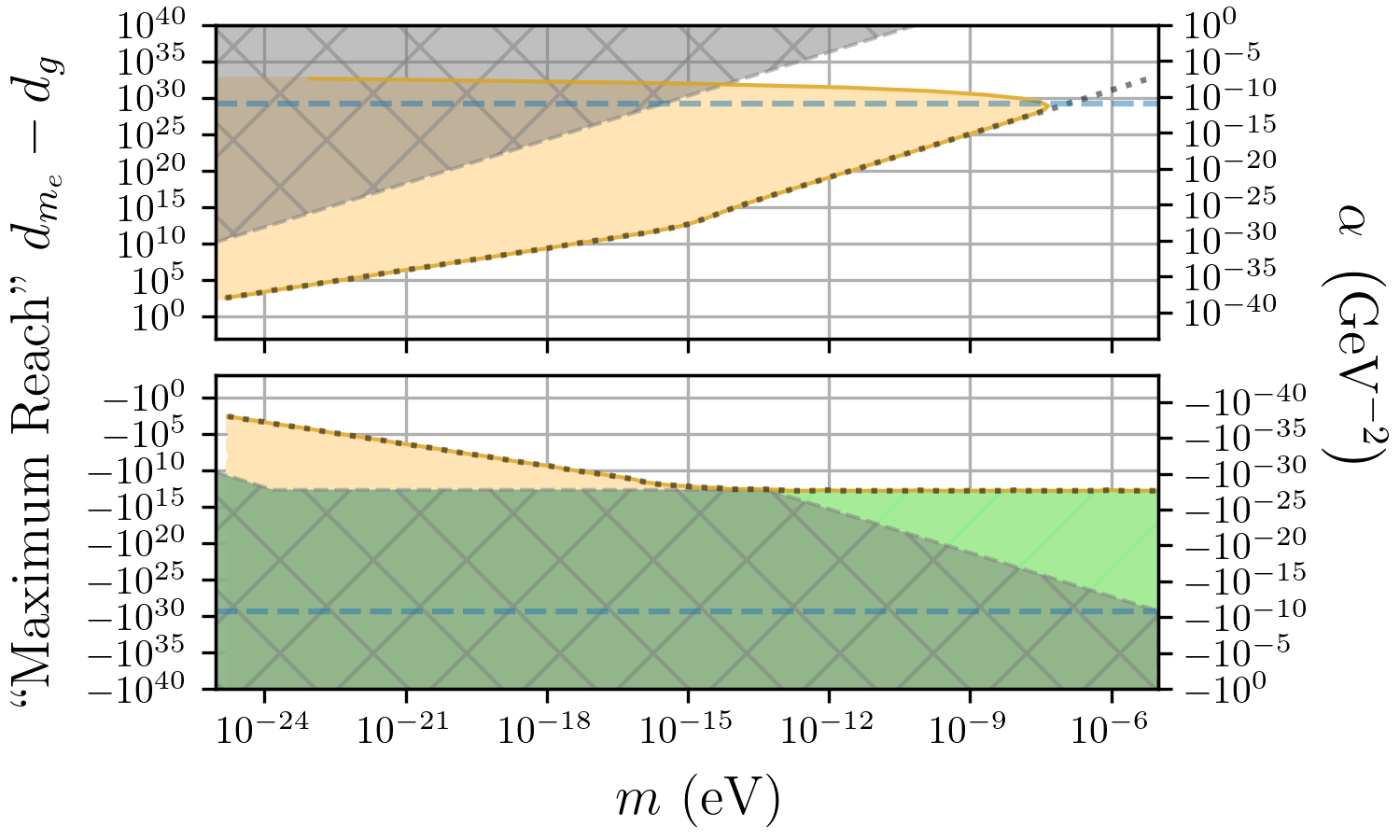}}
\caption{An example of possible modifications to existing  constraints on ultra-light quadratically coupled dark matter. We reproduce the  constraints of Ref.~\cite{Hees:2018fpg} (\textit{black dotted line}) derived from tests of the universality of free-fall, and atomic clock measurements. 
Constrains are plotted in terms of the scalar mass $m$, and the coupling strength in terms of the ``maximum reach'' dilaton parameters on the left vertical axis.  On the right vertical axis we show the equivilent values for our universal coupling  $\alpha$,  calculated using dilaton charges $Q_i$ for iron (see Sec.~\ref{sec:Dilaton_Coefficients}, and for greater detail Refs.~\cite{Damour:2010rp, Damour:2010rm}). These reference values may experience an order of magnitude variation depending on the material chosen.
\textit{Orange} regions show the updated parameter space constraints if each experiment was contained inside a spherical cavity  with the same properties as described in Fig.~\ref{fig:Spherical_Cavity_Interior_Coefficients}.   This is only illustrative of the effects of a cavity on the constraints, and full reevaluation of the constraints requires careful modelling of each experimental environment. \textit{Dark hatched} regions depict where the quadratic model of interactions breaks down: where $|\alpha\phi^2|\gtrsim1$ and, in the case $\alpha<0$, where field solutions around the Earth are expected to be unstable. In \textit{light green} are regions unconstrained by Ref.~\cite{Hees:2018fpg} for the same reason.  Finally, the \textit{blue dashed lines} approximately correspond to $|k|=k_c$ for the cavity solutions, and hence where the cavity enters the regime of strong coupling to the field.}
\label{fig:Hees_Plots}
\end{figure}

In this Section we exemplify  how the  effects of a cavity, described above, could modify existing constraints on quadratically coupled scalars. Specifically we refer to the constraints presented in  Figure 4 of Ref. \cite{Hees:2018fpg}. These constraints derive from tests of the universality of free-fall and atomic clock data \cite{Hees:2018fpg,Banerjee:2022sqg}, and are placed on the dilaton coefficients described in Sec.~\ref{sec:Dilaton_Coefficients}, and which are related to our parameter $\alpha$ via Eqs.~\eqref{eq:Alpha_Definition} and \eqref{eq:Alpha_Conversion}. In particular, the authors of Ref.~\cite{Hees:2018fpg}  use ``maximum reach analysis'', in which it is assumed that all but one term in Eq.~\eqref{eq:Alpha_Definition} is vanishing. This allows for order of magnitude constraints to be placed on the remaining term. 
\\\\
Precise modelling of each experiment is beyond the scope of this work. Here we consider a hypothetical, but illustrative, scenario in which each experiment is contained within the  spherically symmetric cavity discussed in Sec.~\ref{sec:Spherical_Cavity_Profile}, with inner radius $R_1=10\mbox{~cm}$, outer radius $R_2=11\mbox{~cm}$, and density $\rho=2700 \mbox{~kg m}^{-3}$. The revised constraints are shown, orange shaded region,  in Fig.~\ref{fig:Hees_Plots}, as well as the original constraints, dotted line, 
and regions of validity of the quadratic model, outside the grey hashed region. These regions of validity are determined by requiring: $|d_e\kappa^2\phi^2|,~|(d_{\hat{m}}-d_g)\kappa^2\phi^2|, ~|(d_{m_e}-d_g)\kappa^2\phi^2|\ll1 $, for panels (a), (b) and (c) respectively, analogous to the condition $|\alpha\phi^2|\ll 1$.
\\\\
From Fig.~\ref{fig:Hees_Plots}, we clearly observe the impact of the cavity in the $\alpha>0$ (or: $d_e,~(d_{\hat{m}}-d_g),~(d_{m_e}-d_g)>0$) case for strong couplings. In this regime, we observe significant departure from original constraints, rendering large regions of parameter space potentially unconstrainable in a laboratory or satellite experiment. Additionally, whilst the boundary of the  original constraints lies well outside the  region in which the quadratic model of interactions is invalid, cavity effects cause our re-mapped values to intersect with this region. Constraints here cannot be developed without a more comprehensive model of scalar interactions, which is beyond the scope of this work.
\\\\
For  $\alpha<0$ (or: $d_e,~(d_{\hat{m}}-d_g), ~(d_{m_e}-d_g)<0$), the authors of Ref.~\cite{Hees:2018fpg} exclude a large region of parameter space above the first divergence in $\varphi_\oplus$, on the basis of the breakdown of the quadratic model due to divergences in field amplitude. Given that the solution for the field profile around the Earth (Eq.~\eqref{eq:Solution_Uniform_Sphere}) diverges only for specific values of $\alpha$,  marking the full region as unconstrainable may be a more conservative approach than is needed, although  simulating the behaviour of the field in this region is expected to be challenging.
Shown in Fig.~\ref{fig:Hees_Plots} is our own attempt at plotting a region of validity for the quadratic model, incorporating both  conditions on the stability of solutions, and the requirement $|\alpha|\phi^2<1$ for stable solutions. Our region is ultimately identical for $m<\sqrt{\rho_\oplus|\alpha|}$, but is less exclusive for more massive fields. Modifications to any constraints in the $\alpha<0$ panels of Fig.~\ref{fig:Hees_Plots} case, arising from cavity effects, are expected to appear in the region below the blue dashed lines - marking approximately where $|\alpha|=k_c^2/\rho_c$.
\\\\

\section{Differential 5$^{\rm{th}}$ Forces Acting on Cavities}\label{sec:Forces}
We have so far considered cavities as an experimental environment, focussing on the enhancement and suppression of the field within the cavity, and the potential impact this might have on measurements therein. We now consider the evaluation of $5^{\rm th}$ forces acting on spherical cavities themselves. In particular, we ask if changes in the geometry of the cavity, keeping the mass and external radius fixed, leads to detectable differences in the force on the cavities. This is motivated by the fact that test masses of similar geometry have been used in  laboratory based $5^{\rm th}$ force searches; for example the E\"ot-Wash experiment, which uses approximately spherical test bodies with identical mass and external dimension, but differing composition \cite{Schlamminger:2007ht, Wagner:2012ui}. The MICROSCOPE \cite{Touboul:2017grn, MICROSCOPE:2019jix} satellite experiment additionally looks for anomalous differential acceleration between cylindrical Titanium and Platinum bodies towards the Earth. Data from both have been used to set constraints on the properties of scalar field dark matter \cite{Hees:2018fpg}. There are also proposals for an experiment to measure the gravitational force on a Casimir cavity \cite{Archimedes:2014fid}. However, the situation we consider is significantly simplified compared to the experimental environment that would be required for a measurement.
\\\\
In this section, we discuss the $5^{\rm th}$ force experienced by a cavity in an environment where the background scalar field profile can be approximated as varying linearly. We will use this to estimate the difference in force experienced by two cavities in the vicinity of the Earth. To isolate effects due to the structure of the cavity from those due to changes in their size and weight, and to allow differentiation of the $5^{\rm th}$ force from the Newtonian gravitational force on the cavities, we choose the cavities to have the same  mass and external dimension, but different internal structure. In contrast to the constraints discussed in Section \ref{sec:Existing_Constraints}, which assumed ``maximum reach'' values of the dilaton parameters, we forecast the constraints that could be achieved on theories with a universally  coupled scalar.

\subsection{Force on a Single Cavity}
\label{sec:Single_Cavity_Force}
In a non-relativistic setting, the total $5^{\rm th}$ force acting on a cavity with constant density and uniform composition, contained within a volume $V$, is:
\begin{equation}
    \vec{F}_5 = -\frac{1}{2}\alpha\rho_c\int_{V} d^3x~  \nabla\phi^2\;.
    \label{eq:Force_Integral}
\end{equation}
A more detailed description of the derivation of this force is provided in Appendix \ref{app:Force}. Inserting  the field profile around a cavity in an environment with a background scalar gradient, given in  Eq.~\eqref{eq:Solution_Cavity}, we find the total 5th  force acting on a single cavity to be:
\begin{equation}
    \vec{F}_5=-\frac{4\pi \rho_c\alpha}{3}\left[R_2^2\left(a_0+\frac{B^{out}_0}{R_2}\right)\left(a_1R_2+\frac{B^{out}_1}{R_2^2}\right)-R_1^3A^{in}_0A^{in}_1\right]\phi_H^2~\hat{z} \;,
    \label{eq:5th_Force_From_Coefficients}
\end{equation}
where $\hat{z}$ is a unit vector pointing along the direction of the gradient in the linear background field. We compare this to the force on a point particle of the same mass, in the same background:\footnote{We note that this solution can be equally obtained from Eq.~\eqref{eq:5th_Force_From_Coefficients} by first, setting $R_1=0$, and then taking the limit $R_2\rightarrow 0$, whilst fixing the mass of the cavity $M_c=\frac{4\pi}{3}\rho_cR_2^3$. }  
\begin{equation}
    \vec{F}_5^{point} = -\alpha M_ca_0a_1\phi_H^2 ~.
    \label{eq:Point_Mass_Force}
\end{equation}
This point mass approximation gives the false impression that two test bodies with identical masses and values for $\alpha$ will experience identical forces. The  non-trivial dependence of Eq.~\eqref{eq:5th_Force_From_Coefficients} on $R_1,~R_2$, shows that this is not necessarily the case. 
In particular, the point mass approximation fails to accurately describe the force on a body in the strongly coupled limit. The force in
Eq.~\eqref{eq:5th_Force_From_Coefficients} tends asymptotically to a constant limit as  $|\alpha|$ is increased: 
\begin{equation}
  \lim_{\alpha\rightarrow\infty}\vec{F_5}= -4\pi R_2\phi_H^2a_0a_1\hat{z}\;.  
  \label{eq:5th_Force_High_Alpha_limit}
\end{equation}
When $\alpha<0$, Eq.~\eqref{eq:5th_Force_High_Alpha_limit}  again describes the general trend of the force, which will be interspersed with divergences deriving from oscillatory terms.
When $\alpha>0$, the constant limit can be seen as a consequence of Eq.~\eqref{eq:5th_Force_From_Coefficients} being the difference between surface terms at the interior radius and exterior radius of the cavity. Whilst the internal field is exponentially suppressed at $\alpha>k_c^2/\rho_c$, the field at its surface is only suppressed as a power law (as can be seen in equations Eqs.~\eqref{eq:Coefficient_B0_Out}, \eqref{eq:Coefficient_B1_Out}). From a physical perspective, the convergence of Eq.~\eqref{eq:5th_Force_High_Alpha_limit} at large $|\alpha|$ can be understood as  a consequence of matter \textit{generally} screening the field in its neighbourhood. Strong couplings, between the scalar field and matter, lead to strong responses of the field to matter distributions. As such, whilst Eq.~\eqref{eq:5th_Force_From_Coefficients} na\"ively appears to scale with coupling strength, it is balanced by the decreasing amplitude of the field around the cavity.
The point mass approximation of the $5^{\rm th}$ force in Eq.~\eqref{eq:Point_Mass_Force} not only falsely predicts the force should scale indefinitely with $|\alpha|$, but also fails to capture  divergent behaviour present for the $\alpha<0$ case, which hass important effect on the sign of the force.
\\\\
In the regime of weak coupling between the cavity and the field, the point mass solution is more successful. For small $|\alpha|$, Eq.~\eqref{eq:5th_Force_From_Coefficients} takes the form:
\begin{equation}
    \label{eq:5th_Force_Small_Alpha_Limit}\lim_{\alpha\rightarrow0}\vec{F_5}=-\frac{4\pi}{3}\phi_H^2a_0a_1\hat{z}\left[(R_2^3-R_1^3)\rho_c\alpha+\mathcal{O}(\alpha^2)\right] \;,
\end{equation}
for which the leading order term is proportional to the mass of the cavity:
\begin{equation}
    M_c=\frac{4\pi}{3}\rho_c\left(R_2^3-R_1^3\right)~.
\end{equation}
Hence, in this regime, the force on the cavity takes the point mass solution at leading order, whilst geometry-specific contributions appear in $\mathcal{O}(\alpha^2)$ correcting terms. However such terms may still give important contributions to observable phenomena, for example in experiments looking at differential forces, as we discuss in the following section. 

\subsection{Differential Force Evaluation}
\begin{figure}
    \centering
    \includegraphics[width=0.6\linewidth]{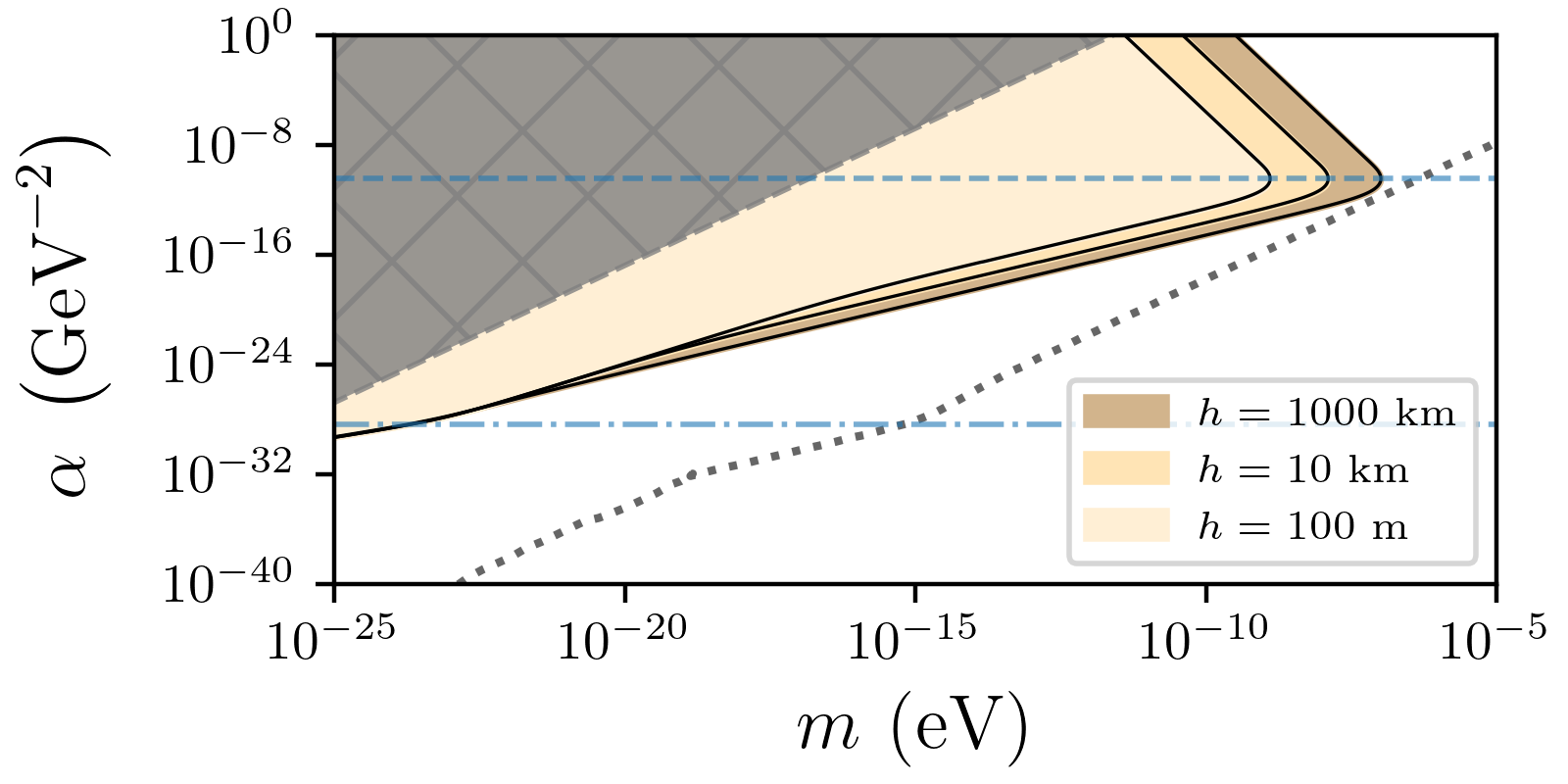}
    \caption{Forecast constraints on the $\alpha>0, m$ parameter space of a light scalar with quadratic couplings to matter, deriving from differential $5^{\rm th}$ forces experienced by cavities in the vicinity of the Earth. The properties of these cavities are drawn from Table \ref{tab:Cavity_Properties}. Constrained regions presume that no signal is discovered up to a sensitivity $\sim100\mbox{ fN}$, akin to current high-precision experiments. Constraints are plotted presuming measurements to be taken at a height $h= 100\mbox{ m}, 10\mbox{ km, and } 1000\mbox{ km}$ above the surface of the Earth. \textit{Blue lines} indicate characteristic values of $\alpha$ for the system: The \textit{dashed line} shows $|\alpha|\sim k_c^2/{\rho_c}$, and the \textit{dot-dashed line} shows $|\alpha|\sim 1/(\rho_\oplus R_\oplus^2)$. For both cavities $1$ and $2$, the value of $k_c$ differs only by a factor of $2$, and hence we only show the line for cavity 1 in this figure. For the  $\alpha>0$ case,  the \textit{black dotted line} gives approximate constraints on $\alpha$, deriving from Ref.~\cite{Hees:2018fpg} under the ``maximal reach'' assumption. As  discussed in Sec.~\ref{sec:Single_Cavity_Force} and Sec.~\ref{sec:Existing_Constraints}, these constraints do not incorporate environmental suppression effects, nor effects deriving from the finite size of test  masses where relevant. The black dotted line should not be used as an indicator of pre-existing constraints for universal $\alpha$. Dark hatched regions indicate where the quadratic model breaks down due to the violation of the condition: $|\alpha\phi^2|\ll1$ at some point in space.}
    \label{fig:Spherical_Cavity_Differential_Forces}
\end{figure}

We now use  Eq.~\eqref{eq:5th_Force_From_Coefficients} to evaluate the differential Earth-cavity $5^{\rm th}$ force on two spherical cavities labelled $1$ and $2$:
\begin{equation}
  \Delta \vec{F_5}=\vec{F_5}^{(1)}-\vec{F_5}^{(2)}\;.
\label{eq:Differential_Force}
\end{equation}
Using the approximate description of the Earth as a uniform spherical body, we determine the coefficients $a_0$ and $a_1$ as in Eq.~\eqref{eq:Earth_Taylor_Expansion}. We make the an illustrative choice to place the cavity at a heights: $h = 100\mbox{ m}, 10\mbox{ km, and } 1000\mbox{ km}$ above the Earth's surface, where in Eq.~\eqref{eq:Earth_Taylor_Expansion} $r_c=R_\oplus+h$. In Sec.~\ref{sec:approx} we discuss further the values of $h$ for which the linearised approximation of field boundary conditions is valid. 
We eliminate the time dependence in  Eq.~\eqref{eq:5th_Force_From_Coefficients},  from the background oscillations in $\phi_H^2$,  by averaging over the oscillations,\footnote{Ponderomotive effects of such oscillations are considered in Ref.~\cite{Zhou:2025wax}.} allowing us to evaluate an order-of-magnitude estimate of the force experienced by the cavity. 
\\\\
Figure~\ref{fig:Spherical_Cavity_Differential_Forces} shows possible constraints imposed by the measurement of the magnitude of the differential $5^{th}$ force to a precision $\sim 100\mbox{ fN}$ using two cavities with the properties shown in Table \ref{tab:Cavity_Properties}, assuming a universal coupling of the scalar to matter. Only forecasts for $\alpha>0$ are presented on account of the regions of instability of our quadratically interacting scalar theory for $\alpha<0$, demonstrated in Fig.~\ref{fig:Hees_Plots} and discussed in Sec.~\ref{sec:sphere}. The figure also shows (in the shaded region), where $|\alpha\phi^2|>1$. In this region of the parameter space we can no longer trust that higher order interactions, in the action of Eq.~\eqref{eq:scalar_field_action}, between the dark matter scalar field and matter are small. For reference, the black dotted line shows the constraints of Ref.~\cite{Hees:2018fpg} on the dilaton coefficient $d_e$, assuming maximum reach, and expressed in terms of the coupling strength $\alpha$ using equations (\ref{eq:Alpha_Definition}) and (\ref{eq:Alpha_Conversion}), as in Figure \ref{fig:Hees_Plots}.\\

\begin{table}
    \centering
    \begin{tabular}{c|c|c|c|c}
                    &  Inner Radius (cm) & Outer Radius  (cm) & Density (kg m$^{-3}$)& Cavity Mass (kg)\\
                    &$R_1$ & $R_2$ & $\rho_c$ &\\\hline
        Cavity 1    & 10 & 11 & 2700 & 3.74  \\
        Cavity 2    & 9 & 11 & 1485 & 3.74
    \end{tabular}
    \caption{Cavity properties used in the generation of Fig. \ref{fig:Spherical_Cavity_Differential_Forces}}
    \label{tab:Cavity_Properties}
\end{table}
\noindent  Fig.~\ref{fig:Spherical_Cavity_Differential_Forces} demonstrates constrained regions of parameter space which initially extend to greater $m$ as $\alpha$ increases. This derives from a differential force which initially grows with $\alpha$. At $\alpha=1/(\rho_\oplus R_\oplus^2)\sim10^{-29}$ GeV$^{-2}$, the trajectory of the boundary of these regions changes subtly - a consequence of $k_\oplus$ now satisfying $|k_\oplus|R_\oplus\gg 1$, and resulting in changes to the background field profile due to the Earth. More significantly, around $\alpha\sim |k^2_c|/{\rho_c}$, which is $ \mathcal{O}\left(10^{-11}\mbox{ GeV}^{-2}\right)$ for both cavities, the boundary of the constrained regions begins to retract to smaller $m$. This is a consequence of the force on each cavity reaching the $\alpha$ independent limit, outlined in 
Eq.~\eqref{eq:5th_Force_High_Alpha_limit}. This limit derives from the finite size of the cavities, and is identical for those in Table~\ref{tab:Cavity_Properties}, due to their identical external radii $R_2$. This effect is not captured in constraints which treat test bodies as point masses.
\\\\
We also observe in Fig.~\ref{fig:Spherical_Cavity_Differential_Forces} that, of the three values of $h$ considered, assuming a height  $h=1000 \mbox{ km}$ constrains the largest region of parameter space. 
Indeed, the $5^{\rm th}$ force in Eq.~\eqref{eq:5th_Force_From_Coefficients},  which depends on $h$ via the product $a_0a_1$, is maximised with respect to $h$ where:
\begin{equation}
    h=R_\oplus\left(-\frac{3}{2}q(k_\oplus R_\oplus)-1\right) \;.
\end{equation}
In  the regime $k_\oplus R_\oplus\gg1$, or $\alpha\gg1/(R^2_\oplus\rho_\oplus)$, the $5^{\rm th}$ force is maximised for  an experiment at $h=R_\oplus/2 \sim 3200 \mbox{ km}$. In the regime $k_\oplus R_\oplus\ll1$, or $\alpha\gg1/(R^2_\oplus\rho_\oplus)$, there is no positive value of $h$ which maximises the $5^{\rm{th}}$ force, and instead  $5^{\rm{th}}$ forces are greater at smaller $h$.
\\\\
We additionally highlight that Fig.~\ref{fig:Spherical_Cavity_Differential_Forces} does not include any suppression of the differential force due the to environment of the experiment, akin to the suppression of forces inside the cavity seen in  Sec.~\ref{sec:Spherical_Cavity_Profile}.  We have assumed that the test cavities are positioned in the vicinity of  the Earth, but with no other material around them. If there is surrounding experimental apparatus, as discussed in Section \ref{sec:Spherical_Cavity_Profile}, we expect a more significant suppression of the field at strong couplings.
\\\\
The significance of the forecasts in Fig.~\ref{fig:Spherical_Cavity_Differential_Forces} depends on assumptions about whether the dark matter scalar couples universally, or preferentially to certain types of matter. If $\alpha$ is allowed to be composition dependent, as per \ref{sec:Dilaton_Coefficients}, we observe that the constraints deriving from purely geometric effects in Fig.~\ref{fig:Spherical_Cavity_Differential_Forces} are largely contained within regions of already constrained parameter space (e.g.~see the black dashed line in the upper panel). Hence, whilst differential forces from geometric effects could be increased by considering larger cavities, we expect that impractical properties would be needed for them to extend existing constraints on models featuring composition dependent couplings.\footnote{Our forecasts are derived assuming universal couplings. If there are compositional dependent couplings then this may lead to additional equivalence principle violation depending on the materials of the two cavities. The described experiment should be a viable means to constrain parameter space if composition dependence is taken into account.}
\\\\
However, we note that constraints such as those in Ref.~\cite{Hees:2018fpg}, illustrated in Fig.~\ref{fig:Hees_Plots}, are {necessarily} predicated on the existence of a composition dependent coupling. They are therefore not applicable to models featuring universal $\alpha$. For such models, Fig.~\ref{fig:Spherical_Cavity_Differential_Forces} demonstrates that an experiment designed to be sensitive to geometrically induced differential $5^{\rm{th}}$ forces could be a viable means to constrain parameter space. ``CubeSats" - with similar form factor and mass to the cavities considered here - may present themselves as a means by which to search for such effects. These present two main advantages. Foremost, they may be placed in high altitude orbits which, as shown in Fig.~\ref{fig:Spherical_Cavity_Differential_Forces}, may improve the sensitivity of the experiment to new parameter space. Secondly, because the satellites are themselves the test masses of the experiment, and are not contained within further apparatus, they do not experience the environmental suppression described in Sec.\ref{sec:Spherical_Cavity_Profile}. This is in contrast to experiments such as MICROSCOPE \cite{Touboul:2017grn,MICROSCOPE:2019jix}, in which the relevant test masses are instead enclosed inside the satellite. An experiment could be realised in which a pair of CubeSats, of similar mass and scale but differing internal structure, could be set into orbit, from which differential deviation from their trajectories could be studied to place bounds on $5^{\rm{th}}$ forces.

\subsection{Discussion of Approximations}
\label{sec:approx}
There are a number of assumptions that we have made in determining the differential force on the cavities, which we justify and discuss here in more detail.
\\\\
First is the assumption that we can approximate the force by its time average in Eq.~\eqref{eq:5th_Force_From_Coefficients}. For an experiment collecting data over time $\tau$, we require $m\tau\gg1$ for this averaging to be valid. Otherwise, the temporal phase of $\phi_H$ is unknown, and hence the order of magnitude Eq.~\eqref{eq:5th_Force_From_Coefficients} is unknown. As a result, only masses that satisfy
\begin{equation}
    m>6.58\times10^{-16}\left(\frac{\mbox{s}}{\tau}\right)~\mbox{eV}\;,
\end{equation}
allow for a robust prediction of the $5^{\rm th}$ force. For  experiments, such as MICROSCOPE and E\"ot-Wash, which operate over time periods $\tau\sim\mathcal{O}(\text{days})$, we require $m\gtrsim10^{-20}\mbox{ eV}$. Clearly, shorter duration experiments will be more restrictive in this regard. For example, if $\tau\sim\mathcal{O}(1~\text{s})$, 
our calculations can only be trusted empirically for $m \gtrsim 10^{-15} ~\mbox{eV}$. 
\\\\
Second, is the assumption that the background field, due to the Earth, is sufficiently slowly varying in the vicinity of the cavity that the linear approximation in Eq.~\eqref{eq:Earth_Taylor_Expansion} is justified.
To ensure this we require the perturbation of the field due to the cavity becomes small compared to the background field before the linear approximation for the background fails. We quantify this by examining the fractional change to each mode $l$ in Eq.~\eqref{eq:Solution_Cavity_Unsimplified}, induced by  the cavity. Assuming the linear approximation, Eq.~\eqref{eq:Earth_Taylor_Expansion}, fails at a length scale  $L$, we require:
\begin{equation}
    \left|\frac{B_0/L}{a_0}\right|\ll 1 ~~~~~~~~~~~~~~~~~~ \left|\frac{B_1/L^2}{a_1L}\right|\ll1 \;,
    \label{eq:Linear_Approximation_Accuracy_Metrics}
\end{equation}
The linearisation of $\varphi_\oplus$ outside of the Earth, about $r_c$, holds for  $r\ll r_c$, motivating $L=r_c$. However, the field profile due to the Earth, may vary significantly at the Earth's surface, due to the rapid change in density, and therefore the effective mass of the scalar. Therefore, our linearisation of the field profile is only valid away from the surface. As such, we choose $L=\min(r_c, ~ h)=h$ where as before, $h$ is the height of the centre of the cavity above the Earth's surface. 
\begin{figure}[t]
    \centering
    \includegraphics[width=0.7\linewidth]{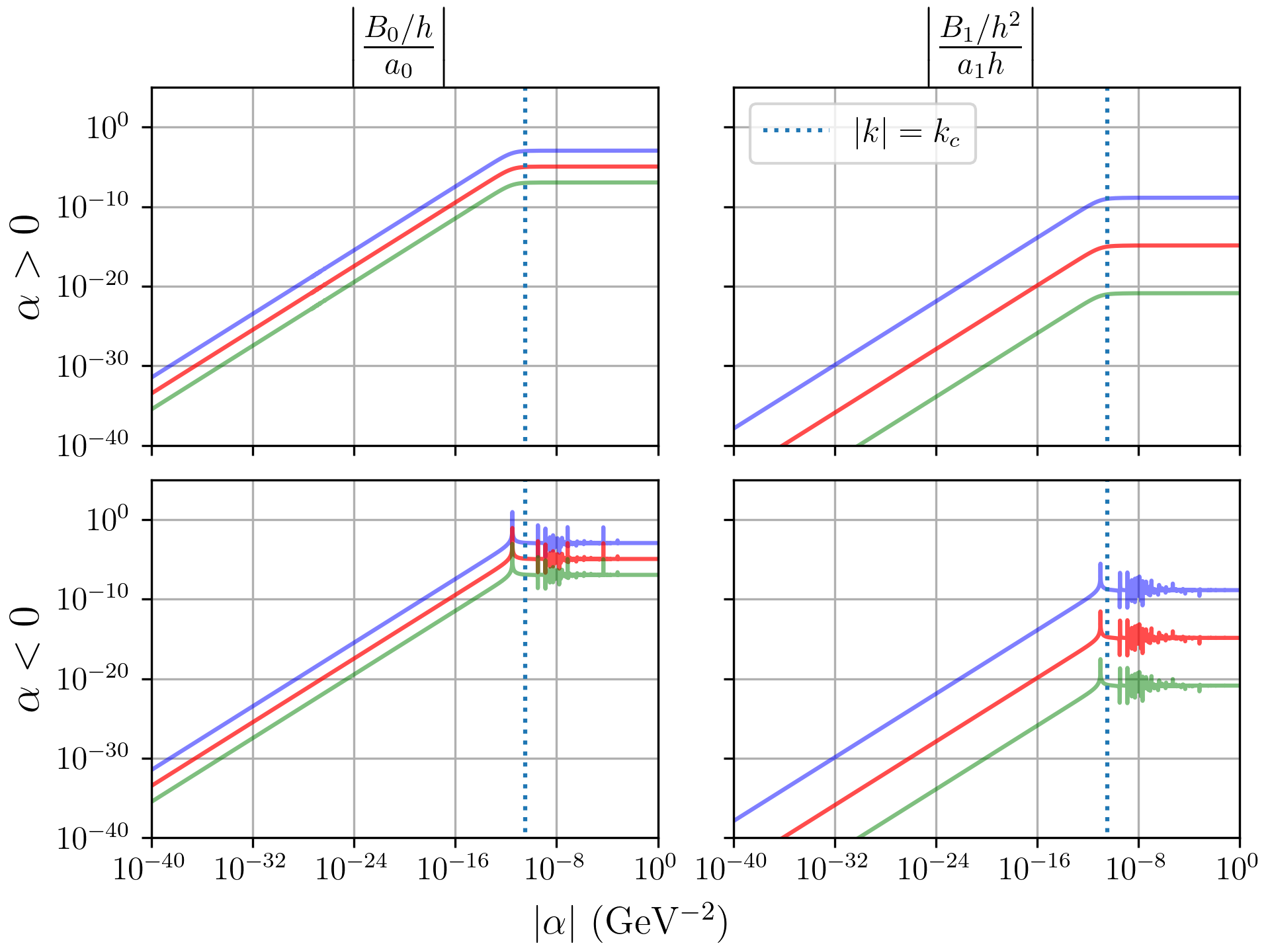}
    \caption{A linearised background field profile is valid as long as the conditions in Eq.~\eqref{eq:Linear_Approximation_Accuracy_Metrics} are satisfied. We plot the relevant values on these conditions for cavity $1$ at $h=100  \mbox{ m},~10 \mbox{ km} ~\mbox{~and}~1000 \mbox{ km}$ (blue, red and green solid lines respectively)
    above the Earth. The left panels show the quantity in Eq.~\eqref{eq:Linear_Approximation_Accuracy_Metrics} corresponding to the condition on the degree $0$ part of the field profile, and the panels on the right to degree $1$ part - displayed as a function of $\alpha$. When these functions are much smaller than one the linearised approximation is valid. 
The top row shows values for $\alpha>0$, and the bottom for $\alpha<0$. 
}
    \label{fig:Lengthscale_Test_Linearisation}
\end{figure}
Figure  \ref{fig:Lengthscale_Test_Linearisation}  shows that the conditions in Eq.~(\ref{eq:Linear_Approximation_Accuracy_Metrics}) generally hold for cavity $1$ at $h=100 ~\mbox{m}$, $10\mbox{ km}$, and  $1000\mbox{ km}$. The approximations may break down due to cavity induced divergences when $\alpha<0$, though some level of enhancement is allowed. The conditions become harder to satisfy at smaller $h$, becoming completely invalid for $h\approx0$. Therefore, for an experiment at the surface of the Earth,  the use of numerical methods will likely be required to fully model the scalar field profile.  
However, we expect that  intra-cavity suppression and enhancement of the scalar field, discussed in Section \ref{sec:Spherical_Cavity_Profile},  should should still be seen.
\\\\
Third, we acknowledge that, we have not accounted for gravitational effects. In particular, an object experiencing a scalar-induced shift to its mass described by Eq.~\eqref{eq:Scalar_Shifted_Atomic_Mass}, is subject to an additional force in a gravitational potential $\Phi$: 
\begin{equation}
    \vec{F}_{5,G}=-\int d^3x~\nabla\Phi \rho(x)\alpha\frac{\phi^2}{2} \;.
    \label{eq:Gravitational_Force}
\end{equation}
Compared against the force computed in Eq.~\eqref{eq:Force_Integral} we expect: 
\begin{equation}
\frac{|\vec{F}_{5,G}|}{ |\vec{F}_{5}|}    \sim\frac{|\nabla\Phi|\phi}{2|\nabla\phi|}\lesssim\frac{g|a_0|}{|a_1|}\;.
\end{equation}
Where $g$ is the classical gravitational acceleration. Hence, we find the gravitational component is only significant for $|\alpha|\lesssim10^{-38}$, which is largely outside of the parameter space considered here. This result is corroborated by numerical evaluation of Eq.~\eqref{eq:Gravitational_Force}, which we compare against Eq.~\eqref{eq:5th_Force_From_Coefficients} to incorporate cavity effects. We do note, however, this assumption may break down in specific regions for $\alpha<0$, where zeros in Eq.~\eqref{eq:q_function}, due to complex $k$, cause the force $\vec{F}_5$ to vanish. Gravitational effects are similarly ignored in Eq.~\eqref{eq:scalar_field_action} on the basis that they are expected to give negligible $\mathcal{O}(\Phi)$ corrections. Further to this, gravitational forces deriving from the energy density of the scalar field around the cavity are expected to scale as $\rho_{DM}/\rho_c$, and hence should be suppressed by a factor $\sim 10^{-24}$ compared to the weight of the cavities in Table~\ref{tab:Cavity_Properties}.
\\\\
Finally, throughout this article we have assumed that all matter distributions, both solid and cavities, are static relative to the dark matter background. The validity of the static approximation for the Earth and for a satellite such as MICROSCOPE were discussed in Ref.~\cite{Burrage:2024mxn}, where it was argued that, after a disturbance, the scalar field approaches its static profile as 
\begin{equation}
    \phi(\vec{x},t) \approx \phi_{\rm st}(r,t) \left(1+ \frac{m(kR-\tanh(kR))}{k}\sqrt{\frac{2}{\pi m t}}\right)\;,
    \label{eq:time_dependence}
\end{equation}
where the `static' profile $\phi_{\rm st}$ is the response to a static object embedded in the time-dependent cosmological background.  It can be shown that the  non-trivial $m$ and $\alpha$ dependence in Eq.~(\ref{eq:time_dependence}) mean that, in the parameter space of interest, the field is always able to respond on the timescales of the motion of matter sources relevant to the experiment ~\cite{Burrage:2024mxn}. Although these results were derived in the context of an instantaneously appearing solid sphere, we expect the same scaling of the response of the field with time to apply for the matter distributions we consider in this work. By a change of frame, we also expect Eq.~(\ref{eq:time_dependence}) to indicate the relaxation time for a static source in a time-dependent dark matter background, for example in the presence of a constant dark matter wind. We note that, in Refs.~\cite{Banerjee:2025dlo,delCastillo:2025rbr}, it was shown that dark matter plane waves impinging on an object from infinity can remove the divergences seen in the $\alpha <0$ case. 

\section{Conclusions}
Quadratically coupled, ultra-light scalar fields are a well-motivated dark matter candidate. Their phenomenology has distinct differences to that of linearly coupled ultra-light scalars, largely driven by a density dependent effective mass for the scalar. In this work we have shown that hollow enclosures, or cavities, can suppress the amplitude of scalar oscillations within them, and that this suppression is particularly effective at strong couplings. We find this effect for both spherical and cylindrical cavities, and for positive and negative choices of coupling. In particular, the suppression is exponential in the case of positive couplings and follows a power law trend for negative couplings. In the case of negative couplings, we also find that the field amplitude may diverge for discrete choices of coupling constant, which depend on the extent of the matter distribution. Such divergences have been observed previously for solid sources. Enhancement of the scalar field is predicted when the coupling constant is in the neighbourhood of these discrete values. However, where the expected enhancement is too great, knowledge of the UV completion of the scalar theory is required to resolve the true behaviour of the field. The general trends of suppression in both amplitude and gradient of the scalar profile makes it challenging to detect the scalar field, and means that existing constraints at strong coupling should be reevaluated. Precise reevaluation likely requires detailed numerical modelling of experimental apparatus, and is beyond the scope of this work, however we illustrate the way in which existing constraints could be modified in Figure \ref{fig:Hees_Plots}. 

We further demonstrate the importance of not treating test masses or sensors as simple point particles, by considering the differential force on two spherical cavities of the same mass and external dimensions, but with different internal radii. We demonstrate that the point particle description fails at strong couplings. Whilst geometric effects are unlikely to significantly extend constraints for models which feature composition dependent couplings between the scalar field and matter, they may be useful for constraining the parameter space of scalar fields which couple universally to matter. An experiment monitoring the motion of CubeSats, or similar devices, might be used to search for these effects.

\begin{acknowledgments}
We would like to thank Benjamin Elder, Yeray Garcia del Castillo, and Joerg Jaeckel for interesting discussions and
collaboration on related topics.
 CB and ET are supported by  STFC Consolidated Grant [Grant No. ST/T000732/1]. AM is supported by EPSRC Quantum Technologies Doctoral Training Partnership (DTP) 2024-25. MR is supported by funding from the U.S. National Science Foundation under Awards PHY-1912380 and PHY-1912514. GR is supported by U.S. Department of Energy Grant No. DE-SC0011665. For open access purposes, the author has applied a CC BY public copyright license to any author accepted manuscript version arising from this submission.
\end{acknowledgments}

\nocite{Macdonald_Code_For_Quadratic_2026}
\newpage
\bibliography{refs}
\newpage
\appendix
\section*{Appendices}
\section{Spherical Cavity Analytic Field Coefficients} \label{app:Coeffs}
The general solutions to Eq.~\eqref{eq:Cavity_Spatial_Equation_Of_Motion} can be expressed as a decomposition in $m=0$ spherical harmonics:
\begin{equation}
        \varphi_c(r) = \begin{cases}
        \sum_{l=0}^\infty \left(A_l^{in}r^l + B_l^{in}r^{-l-1}\right)P_l^0(\cos\theta) & r<R_1\\
        \sum_{l=0}^\infty \left(\begin{split}
            &A_l^{walls}[(-i)^lj_l(-ikr)] \\+&~ B_l^{walls}[(-i)^{-l-1} y_{l}(-ikr)]
        \end{split}\right)P_l^0(\cos\theta) & R_1<r<R_2\\
        \sum_{l=0}^\infty \left(A_l^{out}r^l + B_l^{out}r^{-l-1}\right)P_l^0(\cos\theta) & r>R_2
    \end{cases}
    \label{eq:Solution_Cavity_Unsimplified}
\end{equation}
where $A_l$ and $B_l$ are constants of integration,
$k=\sqrt{\rho_c\alpha}$ and $i$ is the imaginary unit. 
$P^0_l(\cos\theta)$ are 0th order Legendre polynomials of $l^{th}$ degree, and $j_l(x),~ y_l(x)$ are Spherical Bessel functions of the first and second kind respectively. The choice to include the factors $(-i)^{p}$ in front of the Bessel functions ensures that  the coefficients $A_l^{walls}$, $B_l^{walls}$ are real for $k\in\mathbb{R}$. A valid solution requires continuity and differentiability at $R_1$ and $R_2$, regularity at the origin, and that the field approaches the background solution far from the cavity. 
Amalgamating these conditions, we ascribe the following expressions to the constants, noting that all but the degree $0$ and degree $1$ terms vanish:

\begin{alignat}{2}
    &A^{in}_0=&&\frac{a_0}{kR_1\sinh(k(R_2-R_1))+\cosh(k(R_2-R_1))}
    \\
    \notag\\
    &A^{in}_1=&&\frac{3a_1 kR_2}{3\sinh(k(R_2-R_1))+3kR_1\cosh(k(R_2-R_1))+(kR_1)^2\sinh(k(R_2-R_1))}\label{eq:A0_in}
    \\
    \notag\\
    &A^{walls}_0=&&\frac{a_0(\cosh(kR_1)-kR_1\sinh(kR_1))}{kR_1\sinh(k(R_2-R_1))+\cosh(k(R_2-R_1))}
    \\
    \notag\\
    &B^{walls}_{0}=&&\frac{a_0(kR_1\cosh(kR_1)-\sinh(kR_1))}{kR_1\sinh(k(R_2-R_1))+\cosh(k(R_2-R_1))}
    \\
    \notag\\
    &A^{walls}_1 =&& \frac{-3a_1 R_2(3\cosh(kR_1)-3kR_1\sinh(kR_1)+(kR_1)^2\cosh(kR_1))}{3\sinh(k(R_2-R_1))+3kR_1\cosh(k(R_2-R_1))+(kR_1)^2\sinh(k(R_2-R_1))}
    \\
    \notag\\
    &B^{walls}_1 =&& \frac{-3a_1 R_2(3\sinh(kR_1)-3kR_1\cosh(kR_1)+(kR_1)^2\sinh(kR_1))}{3\sinh(k(R_2-R_1))+3kR_1\cosh(k(R_2-R_1))+(kR_1)^2\sinh(k(R_2-R_1))}
    \\
    \notag\\
    &B_0^{out}=&&\frac{a_0}{k}\frac{\sinh(k(R_2-R_1))-k(R_2-R_1)\cosh(k(R_2-R_1))-k^2R_1R_2\sinh(k(R_2-R_1))}{kR_1\sinh(k(R_2-R_1))+\cosh(k(R_2-R_1))}\label{eq:Coefficient_B0_Out}
    \\
    \notag\\
    &B_1^{out}=&&\frac{-a_1R_2}{k^2}\dfrac{\left(\splitdfrac{\sinh(k(R_2-R_1))(9+3k^2(R_2^2+R_1^2)+k^4R_1^2R_2^2-9k^2R_1R_2)}{+ \cosh(k(R_2-R_1))(3k^3R_1R_2(R_2-R_1)-9k(R_2-R_1))}\right)}{3\sinh(k(R_2-R_1))+3kR_1\cosh(k(R_2-R_1))+(kR_1)^2\sinh(k(R_2-R_1))} \label{eq:Coefficient_B1_Out}
\end{alignat}
where $a_0$ and $a_1$ are defined by the linear boundary conditions assumed in Eq.~\eqref{eq:Cavity_Infinity_Boundary_Condition}.

\section{Solution for an infinitely long cylindrical cavity} \label{app:infinite_cyle}
To determine the field profile around an infinitely long cylindrical cavity, with internal radius $R_1$, external radius $R_2$ and density $\rho_c$, we solve the equation
\begin{equation}
\partial_r^2\varphi(r) + \frac{1}{r}\partial_r\varphi(r) = \alpha \rho(r) \varphi(r)\enspace,
\end{equation}
with
\begin{equation}
    \rho(r) = \rho_c\, \Theta(r-R_1)\Theta(R_2 -r)\enspace.
\end{equation}
where $\Theta(x)$ is the Heaviside step function.

The solution is
\begin{align*}
\Phi_{\text{I}}(r) &= c_1 + c_2 \log(r) & r<R_1\\
\Phi_{\text{II}}(r) &=
c_3 J_0\left(i k r\right) 
+
c_4 Y_0\left(-i kr\right)  & R_1\leq r \leq R_2\\
\Phi_{\text{III}}(r) &= c_5 + c_6 \log(r) & r > R_2\enspace,
\end{align*}
where $J_\nu$ and $Y_\nu$ are Bessel function of the first and second kind, respectively.

We impose that the solution be finite at $r=0$, its first derivative be continuous at $r=R_2$ and $r=R_2$, and $\varphi(r) = 1$ at $r=R_{\text{max}}$. In this case, setting the boundary condition at $r\to\infty$ is unphysical, since, for any real-world cylinder, there will always be a finite distance from the cylinder from which it will be possible to ``see" its finite length. In Section~\ref{sec:Cylinder}, we have set $R_{\text{max}}$ equal to the length of the simulation box in the $r$-direction.

Defining the dimensionless lengths $\tilde{R_i} = kR_i$, the coefficients have then the following expressions
\begin{eqnarray}
c_1 &=& \left(\tilde{R_1} \bigg[
    K_1(\tilde{R_1})
    \left(
        I_0(\tilde{R_2})
        - \tilde{R_2} \, I_1(\tilde{R_2}) \log\left(\frac{R_2}{R_{\text{max}}}\right)
    \right) \right.\\ \nonumber
    &&\left.+ I_1(\tilde{R_1})
    \left(
        K_0(\tilde{R_2})
        + \tilde{R_2} \, K_1(\tilde{R_2}) \log\left(\frac{R_2}{R_{\text{max}}}\right)
    \right)
\bigg] \right)^{-1}\\
c_2 &=& 0\\
c_3 &=&\frac{i}{2} \pi \tilde{R_1} \, Y_1(-i \tilde{R_1}) \, c_1 \\
c_4 &=&  -\frac{1}{2}\pi \tilde{R_1} \, I_1(\tilde{R_1}) \, c_1 \\
c_5 &=& \tilde{R_1}  \bigg( 
    K_1(\tilde{R_1}) \left( I_0(\tilde{R_2}) - \tilde{R_2} I_1(\tilde{R_2}) \log R_2 \right) 
    + I_1(\tilde{R_1}) \left( K_0(\tilde{R_2}) + \tilde{R_2} K_1(\tilde{R_2}) \log(R_2) \right) 
\bigg) c_1 \\[1ex]
c_6 &=&  \tilde{R_1}\tilde{R_2} \left( I_1(\tilde{R_2}) K_1(\tilde{R_1}) - I_1(\tilde{R_1}) K_1(\tilde{R_2}) \right) c_1
\end{eqnarray}
where $I_\nu$ and  $K_\nu$ are modified Bessel functions of the first and second kind, respectively.

\section{Cavity $5^{\rm{th}}$ Forces} \label{app:Force}
The classical contribution to the $5^{\rm th}$ force experienced by a massive object in the presence of a scalar field can be evaluated by either of two means. Most simply, if the object is assumed to be irrotational and non-relativistic, Eq. \eqref{eq:scalar_field_action} may be modified by making the replacement: $\rho(x)\rightarrow \rho(\vec{x}-\vec{X})$ and inserting a kinetic term into the action \[\frac{1}{2}\rho(x)\dot{X}^2\] where $\vec{X}(t)$ is the centre of mass of the object, or some similar point. Deriving the equation of motion for $\vec{X}$ determines the net force on the object. Additionally, one can determine the non-conserved part of the scalar field stress-energy tensor, which details the exchange of momentum between the field and matter to which it is coupled:
\begin{equation}
    F_5^\nu=\int d^3x ~\nabla_\mu T^{\mu\nu}_\phi
\end{equation}
Both ultimately give the same expression:
\begin{equation}
    \vec{F_5}=\int d^3x~\nabla\left(\rho(x)\alpha\right)\frac{\phi^2}{2}=-\int d^3x~\rho(x)\alpha\frac{\nabla\phi^2}{2}
\end{equation}
Where integrals are given over all space, and the absence of a matter density at infinity means surface terms may be dropped. For a spherical cavity, this simplifies to give
Eq. \eqref{eq:Force_Integral}, which can then be evaluated as the difference of two volume integrals, covering uniform spheres of radius $R_1$ and $R_2$:
\begin{equation}
    \vec{F}_5 = -\frac{1}{2}\alpha\rho_c\left[\int_{R_2} d^3x~  \nabla\phi^2
 - \int_{R_1} d^3x~  \nabla\phi^2
\right]    
\end{equation}
Which may be further simplified through application of divergence theorem
\begin{equation}
    \vec{F}_5 = -\frac{1}{2}\alpha\rho_c\left[\int_{R_2} d^2 \Omega  ~R_2^2~\hat{r}~  \phi^2
 - \int_{R_1} d^2 \Omega  ~R_1^2~\hat{r}~  \phi^2
\right]
\end{equation}
Due to the cylindrical symmetry of the problem, integrating over the azimuthal angular coordinate discards components orthogonal to the afore-defined $\hat{z}$ direction, giving:
\begin{equation}
\begin{split}
    \vec{F}_5 = -\pi\alpha\rho_c&\left[\int_0^\pi d\theta\sin\theta ~R_2^2~(\hat{z}\cos\theta)~  \left(a_0+\frac{B^{out}_0}{R_2}+\left(a_1R_2+\frac{B^{out}_1}{R_2^2}\right)\cos\theta\right)^2 \right. \\ &\left.
 - \int_0^\pi d\theta \sin\theta  ~R_1^2~(\hat{z}\cos\theta)~  (A^{in}_0+A^{in}_1R_1\cos\theta)^2
\right]
\end{split}
\end{equation}
From which only the cross terms between angular modes survive in each integral, yielding Eq. \eqref{eq:5th_Force_From_Coefficients}. Substitution of coefficients yields an analytic expression:
\begin{equation}
    \vec{F}_5=\frac{-4\pi \phi_H^2a_0a_1\hat{z}}{k}\frac{k(R_2-R_1)\cosh(k(R_2-R_1))+(k^2R_2R_1-1)\sinh(k(R_2-R_1))}{\cosh(k(R_2-R_1))+kR_1\sinh(k(R_2-R_1)))}
\end{equation}
\newpage

\end{document}